\documentclass[fleqn,usenatbib]{mnras}

\let\VANthebibliography\thebibliography
\def\thebibliography{\DeclareRobustCommand{\VAN}[3]{##3}\VANthebibliography}

\usepackage[T1]{fontenc}
\usepackage{graphicx}
\usepackage{amsmath}
\usepackage{amssymb}
\usepackage{caption}
\usepackage{subcaption}
\usepackage{ae,aecompl}
\usepackage{multirow,array}
\usepackage{adjustbox}
\usepackage{color,soul}
\usepackage[dvipsnames]{xcolor}
\usepackage{float}
\usepackage{datetime}
\usepackage{ulem}
\usepackage{newtxtext,newtxmath}

\renewcommand{\theenumi}{(\arabic{enumi})}

\title[Redshift evolution of galaxy subpopulations]{Synergies between low- and intermediate-redshift galaxy populations revealed with unsupervised machine learning}

\author[Sebastian Turner et al.]{Sebastian Turner$^{1,}$\thanks{E-mail: s.turner1@2012.ljmu.ac.uk}, Małgorzata Siudek$^{2,3,}$\thanks{E-mail: msiudek@ifae.es}, Samir Salim$^{4}$, Ivan K. Baldry$^{1}$, Agnieszka Pollo$^{3,5}$,\newauthor Steven N. Longmore$^{1}$, Katarzyna Małek$^{3,6}$, Chris A. Collins$^{1}$, Paulo J. Lisboa$^{7}$, Janusz Krywult$^{8}$,\newauthor Thibaud Moutard$^{9}$, Daniela Vergani$^{10}$, and Alexander Fritz$^{11}$\\
$^{1}$Astrophysics Research Institute, Liverpool John Moores University, 146 Brownlow Hill, Liverpool, L3 5RF, UK\\
$^{2}$Institut de F’isica d’Altes Energies, The Barcelona Institute of Science and Technology, 08193 Bellaterra, Spain\\
$^{3}$National Centre for Nuclear Research, ul. Hoza 69, 00-681 Warsaw, Poland\\
$^{4}$Department of Astronomy, Indiana University, Bloomington, IN 47405, USA\\
$^{5}$Astronomical Observatory of the Jagiellonian University, ul. Orla 171, 30-244 Krak\'ow, Poland\\
$^{6}$Aix Marseille Univ. CNRS, CNES, LAM Marseille, France\\
$^{7}$Department of Applied Mathematics, Liverpool John Moores University, Byrom Street, Liverpool, L3 3AF, UK\\
$^{8}$Institute of Physics, Jan Kochanowski University, ul. Swietokrzyska 15, 25-406, Kielce, Poland\\
$^{9}$Department of Astronomy \& Physics and the Institute for Computational Astrophysics, Saint Mary’s University, 923 Robie Street, Halifax,\\ Nova Scotia, B3H 3C3, Canada\\
$^{10}$INAF - OAS Bologna, Via P. Gobetti 93, I-40129 Bologna, Italy\\
$^{11}$Lichtenbergstraße 8, D-85748 Garching, Germany\\}

\date{Accepted XXX. Received YYY; in original form ZZZ}
\pubyear{\the\year}

\begin{document}
\label{firstpage}
\pagerange{\pageref{firstpage}--\pageref{lastpage}}
\maketitle

\begin{abstract}
The colour bimodality of galaxies provides an empirical basis for theories of galaxy evolution. However, the balance of processes that begets this bimodality has not yet been constrained. A more detailed view of the galaxy population is needed, which we achieve in this paper by using unsupervised machine learning to combine multi-dimensional data at two different epochs. We aim to understand the cosmic evolution of galaxy subpopulations by uncovering substructures within the colour bimodality. We choose a clustering algorithm that models clusters using only the most discriminative data available, and apply it to two galaxy samples: one from the second edition of the GALEX-SDSS-WISE Legacy Catalogue (GSWLC-2; $z \sim 0.06$), and the other from the VIMOS Public Extragalactic Redshift Survey (VIPERS; $z \sim 0.65$). We cluster within a nine-dimensional feature space defined purely by rest-frame ultraviolet-through-near-infrared colours. Both samples are similarly partitioned into seven clusters, breaking down into four of mostly star-forming galaxies (including the vast majority of green valley galaxies) and three of mostly passive galaxies. The separation between these two families of clusters suggests differences in the evolution of their galaxies, and that these differences are strongly expressed in their colours alone. The samples are closely related, with star-forming/green-valley clusters at both epochs forming morphological sequences, capturing the gradual internally-driven growth of galaxy bulges. At high stellar masses, this growth is linked with quenching. However, it is only in our low-redshift sample that additional, environmental processes appear to be involved in the evolution of low-mass passive galaxies.
\end{abstract}

\begin{keywords}
galaxies: general - galaxies: evolution - galaxies: star formation - galaxies: stellar content - galaxies: statistics - methods: statistical
\end{keywords}



\section{Introduction}
\label{sec:intro}

The composition of a galaxy is subject to the influence of an ever-changing balance of astrophysical and cosmological processes acting upon it. Hence, chronicling of the evolutionary history of a galaxy requires a precise knowledge of its present contents. A galaxy expresses its contents (stars, gas, dust, etc.) in its spectral energy distribution (SED). Therefore, inventory of the composition of a galaxy requires measurement of the radiation that it emits as a function of wavelength \citep{CONROY-2013}. It is impractical to measure full galaxy spectra that span large wavelength ranges (e.g. ultraviolet-through-infrared), especially for the large number of galaxies needed for a robust statistical study of galaxy evolution. Instead, their SEDs must be inferred from curtailed, summary measurements.

Colours are the simplest such measurements. Optical colours have been used to probe the contents of galaxies since the infancy of extragalactic astrophysics \citep{ROBERTS+1963}. Early studies matched sums of individual stellar spectra (i.e. synthetic composite spectra) to the observed optical colours of galaxies in order to discern their stellar content (e.g. \citealt{SPINRAD-1962,SPINRAD+1971,FABER-1972}). This method was superseded by stellar population synthesis (SPS), which uses theoretical models of stellar evolution to set astrophysical constraints upon these synthetic composite spectra (e.g. \citealt{BRUZUAL+2003,MARASTON+2005}). The advancement of the scope of SPS out to ultraviolet wavelengths and the incorporation of infrared emission models has facilitated the estimation of the full ultraviolet-through-infrared SEDs of galaxies from their observed colours (e.g. \citealt{ILBERT+2006,DACUNHA+2008,BOQUIEN+2019}). SEDs spanning these wavelength regimes are governed in their shapes chiefly by stellar emission, and by attenuation (in the ultraviolet and optical) and re-emission (in the infrared) of stellar emission by interstellar dust.

The discovery of a bimodality in the two-dimensional optical colour distribution of galaxies \citep{STRATEVA+2001,BALDRY+2004} has begotten a simple empirical paradigm of galaxy evolution. Galaxies generally go from being blue and star-forming to being red and passive. This change in their colours (and quenching of their star formation) is accompanied for the most part by a change in their morphologies from disc- (`late-type') to spheroid-dominated (`early-type') and an increase in their local environmental densities \citep{BALDRY+2006,BAMFORD+2009}. A variety of processes have been proposed as drivers of galaxy evolution (see reviews by \citealt{KORMENDY+2004} and \citealt{BOSELLI+2006}) but their interplay is poorly understood. Furthermore, exceptions to this paradigm \citep{SCHAWINSKI+2009c,MASTERS+2010} complicate the issue. Studies aiming to disentangle the interplay of evolutionary processes have focused on galaxies between the two peaks of the colour bimodality (\citealt{FRITZ+2014,SCHAWINSKI+2014,SMETHURST+2015,MOUTARD+2016b,GU+2018,MANZONI+2019}; Krywult et al. in prep), a region called the `green valley' \citep{MARTIN+2007}. As galaxies under the direct influence of evolutionary processes, they are ideally poised to enable an understanding of how galaxies transition from blue to red.

Bimodalities of galaxies have since also been observed in colours involving ultraviolet and near-infrared magnitudes \citep{WYDER+2007,WILLIAMS+2009,ARNOUTS+2013}. Different colours, though, yield slightly different bimodalities; for example, galaxies occupying the blue peak of the $g-r$ bimodality may instead occupy the green valley of the $NUV-r$ bimodality \citep{SALIM+2014}, because optical-optical colours probe star formation over longer timescales than ultraviolet-optical colours do. Hence it is clear that, for a complete description of the evolution of galaxies in the context of the bimodality and the green valley, several colours spanning the ultraviolet-through-near-infrared wavelength regime must be considered simultaneously. Machine learning techniques, which can parse multiple features at once, are well suited to the task. Exploration of the multi-dimensional ultraviolet-through-near-infrared colour distribution of galaxies may overcome degeneracies that exist in two-dimensional colour distributions, uncover substructures to the established bimodality, and reveal the extent to which the ultraviolet-through-near-infrared colours of galaxies express their evolution and assembly histories.

The adoption of machine learning techniques within astronomy and astrophysics was primarily a response to the enormous data volumes anticipated from forthcoming surveys (e.g. $20$ TB per night from the Legacy Survey of Space and Time; \citealt{IVEZIC+2019}). While fulfilling the demand for automated data analysis methods, these techniques also invite a renewed examination of our understanding of astrophysics due to their ability to distill interpretable models from complex, multi-dimensional input data that may be difficult to fully visualise. Supervised techniques are useful for mapping existing domain knowledge onto new data. A supervised classification algorithm, for example, may assign labels to previously unseen observations after being trained on prelabelled observations. Unsupervised techniques, on the other hand, demonstrate substantial promise for exploration and discovery because they are less reliant on prior knowledge than supervised techniques. An unsupervised clustering algorithm, for example, assigns labels to observations in accordance with their intrinsic similarity to one another (i.e. the distances between observations in terms of the features used to represent them). Unsupervised techniques, then, construct models that are driven purely by the structure of input data, and require no training. They may therefore be said to express the `natural' structure of the input data rather than expressing structures imposed upon it by assumptions that are explicitly built into the use of supervised techniques. The use of unsupervised techniques does, though, incorporate implicit assumptions, and the precise definition of similarity can vary between techniques. Ensuring the astrophysical utility of these models hence requires carefully considered choices of algorithm and features.

A growing literature has emerged in recent years, reporting the results of the application of unsupervised techniques to various astrophysical contexts (see \citealt{BARON+2019} and \citealt{BALL+2010} for comprehensive reviews). Clustering has been used, for example, to partition galaxies on the basis of their pixel data \citep{HOCKING+2017,HOCKING+2018,MARTIN+2020}, their spectra \citep{SANCHEZALMEIDA+2010,DESOUZA+2017}, their SEDs \citep{SIUDEK+2018a,SIUDEK+2018b}, and their derived astrophysical features \citep{BARCHI+2016,TURNER+2019}. Dimensionality reduction, which can extract important or discriminative information from large ensembles of input features, has been used, for example, to produce simplified projections of galaxy samples based on their multi-wavelength photometry \citep{STEINHARDT+2020} and their estimated SEDs \citep{DAVIDZON+2019,HEMMATI+2019}, and to classify their spectra \citep{YIP+2004,MARCHETTI+2013}.

In this paper, we describe work that builds on that of \cite{SIUDEK+2018a,SIUDEK+2018b}. They applied a clustering algorithm to partition galaxies observed by the VIMOS Public Extragalactic Redshift Survey (VIPERS; \citealt{SCODEGGIO+2018}). They chose the Fisher Expectation-Maximisation (\texttt{FEM}) algorithm, which implements a clustering approach called the `Discriminative Latent Mixture' (DLM) model. The algorithm incorporates dimensionality reduction as it iterates rather than as a part of any preparation of the input data ahead of clustering. This ensures that improvements to the estimated parameters of the model are adaptive, and that the clustering uses only the most important information available from the input features. They aimed to establish the ability of \texttt{FEM} to determine a naturally defined, astrophysically meaningful partition in a feature space of high dimensionality (i.e. containing more potentially discriminative information than lower dimensionalities). Their feature space was defined by spectroscopic redshifts and $12$ rest-frame ultraviolet-through-near-infrared colours. The $12$ clusters that they determined revealed substructure to the established colour bimodality of galaxies, distinguishing subpopulations of galaxies that overlapped in two-dimensional colour distributions. In addition, their clusters correlated with a variety of astrophysical features including stellar masses, morphologies, and emission-line strengths.

We adapt the approach of \cite{SIUDEK+2018a,SIUDEK+2018b} to compare samples of galaxies at two different redshifts. Our aim is to use clustering to characterise the structures of the samples in a common feature space of high dimensionality, to examine similarities and differences between these structures at the two cosmic epochs, and to interpret these similarities and differences in the context of theories of galaxy evolution. While each cluster will constitute a class of galaxies that are intrinsically similar to one another, connections between clusters will chart the evolution of galaxies through the feature space. Hence, we also aim to establish how strongly the evolutionary histories of galaxies, which are ordinarily inferred using a combination of various types of features (e.g. photometric, spectroscopic, morphological), are encoded in just their ultraviolet-through-infrared colours. Our sample of galaxies at low redshift ($z \sim 0.06$) is drawn from the second edition of the GALEX-SDSS-WISE Legacy Catalogue (GSWLC-2; \citealt{SALIM+2018}), and our sample of galaxies at intermediate redshift ($z \sim 0.65$) is based on the VIPERS sample of \cite{SIUDEK+2018a}. We prepare our samples carefully to ensure a fair comparison of galaxies from different cosmic epochs and different surveys, and to mitigate methodological influences on the clustering outcomes. We also adjust the input features, defining nine neighbouring rest-frame colours that, together, represent the shapes of the ultraviolet-through-near-infrared SEDs of the galaxies in our samples, and thus enable insight into their evolution.

The remainder of this paper proceeds as follows. In Section \ref{sec:data}, we introduce our samples, the data we use to represent and analyse the galaxies that they contain (including the estimation of their SEDs), and the measures that we take to ensure a fair comparison between them. In Section \ref{sec:meth}, we explain the DLM model and how \texttt{FEM} algorithm implements it, and we describe the feature space within which we cluster our samples. In Section \ref{sec:res}, we present the outcomes of the clustering, and in Section \ref{sec:disc}, we offer our interpretation thereof. Finally, in Section \ref{sec:conc}, we summarise, make concluding statements, and suggest future directions for our work. Where required, we assume a ($H_{0}$, $\Omega_{m}$, $\Omega_{\Lambda}$) = ($70$ km s$^{-1}$ Mpc$^{-1}$, $0.3$, $0.7$) cosmology for our calculations.

\section{Data}
\label{sec:data}

\subsection{GALEX-SDSS-WISE Legacy Catalogue 2}
\label{subsec:gswlc}

The second edition of the GALEX-SDSS-WISE Legacy Catalogue (GSWLC-2; \citealt{SALIM+2016,SALIM+2018}) was assembled using Data Release 10 (DR10; \citealt{AHN+2014}) of the Sloan Digital Sky Survey (SDSS; \citealt{YORK+2000}). GSWLC-2 aimed to characterise the star formation activity and dust content of galaxies in the local Universe. It contains all SDSS DR10 galaxies that meet the following criteria:

\begin{itemize}
    \item have apparent $r$-band petrosian magnitudes $< 18$,
    \item have spectroscopic redshifts within the range $0.01 < z < 0.3$,
    \item lie within the Galaxy Evolution Explorer (GALEX; \citealt{MARTIN+2005, MORRISSEY+2007}) observation footprint, whether they were detected by GALEX or not.
\end{itemize}

The lower redshift limit was imposed to exclude foreground stars, and particularly close galaxies with potentially unreliable photometry and/or distance estimates. Retaining galaxies that were not actually detected by GALEX itself preserves the optical selection of SDSS. In all, these criteria select $659,229$ SDSS DR10 galaxies.

$u$-, $g$-, $r$-, $i$-, and $z$-band optical photometry for galaxies in GSWLC-2 was drawn from SDSS. \texttt{modelMag} magnitudes, which are based on profile fits, were selected due to the accuracy of their colours. These \texttt{modelMag} magnitudes were corrected for extinction due to Milky Way dust using the empirical \cite{YUAN+2013} coefficients.

The SDSS optical photometry was supplemented with near- ($NUV$) and far-ultraviolet ($FUV$) photometry from GALEX's final data release (GR6/7). GALEX conducted surveys at varying depths: an All-sky Imaging Survey (which observed several targets per orbit), a Medium Imaging Survey (one target per orbit), and a Deep Imaging Survey (several orbits per target). These surveys were nested, such that it is possible for a galaxy to have been observed at more than one depth (although an observation of a galaxy at a given depth does not guarantee an observation of the same galaxy at shallower depths). Here we use the UV photometry for galaxies in GSWLC-2 based on the deepest available observation of each galaxy (catalogue GSWLC-X2). \cite{SALIM+2016} applied corrections to mitigate systematic offsets between the SDSS and GALEX photometry, which arose mostly due to the blending of sources in GALEX's low-resolution images. \cite{PEEK+2013} corrections for extinction due to Milky Way dust were applied to the UV photometry. UV photometry in at least one of GALEX's two bands (almost always $NUV$ if just one) is available for $65$ per cent of GSWLC-2 galaxies, and for $80$ per cent of the galaxies in our final GSWLC-2 sample (Section \ref{subsubsec:finalg}).

Wide-field Infrared Survey Explorer (WISE; \citealt{WRIGHT+2010}) observations at $12$ and $22$ $\mu$m (channels W3 and W4 respectively) were used to provide mid-infrared (MIR) photometry for GSWLC-2 galaxies. \cite{SALIM+2018} opted for unWISE \citep{LANG+2016} forced photometry, which was based directly on SDSS source positions and profiles. MIR photometry in at least one of channels W3 and W4 is available for $78$ per cent of GSWLC-2 galaxies, and for $87$ per cent of the galaxies in our final GSWLC-2 sample (Section \ref{subsubsec:finalg}).

\subsubsection{GSWLC-2 rest-frame SEDs}
\label{subsubsec:gseds}

The rest-frame SEDs of GSWLC-2 galaxies were estimated using the Code Investigating GALaxy Emission (\texttt{CIGALE}; \citealt{NOLL+2009,BOQUIEN+2019}). Synthetic spectra generated by \texttt{CIGALE} were validated against the available observed UV-through-optical photometry in order to constrain the SEDs. Details of this fitting procedure are described at length in \cite{SALIM+2016,SALIM+2018}; here, we offer a brief summary.

Synthetic spectra were generated using \cite{BRUZUAL+2003} simple stellar population templates, based on a \cite{CHABRIER-2003} initial mass function and with metallicities of $\log_{10}(Z) = -2.4$, $-2.1$, $-1.7$ ($\sim Z_{\odot}$), or $-1.3$. These templates were combined with Myr-resolution star formation histories (SFHs) consisting of two exponentially declining episodes of star formation, producing an old and a young population. Absorption of stellar emission by dust was implemented via a \cite{NOLL+2009} generalisation of the \cite{CALZETTI+2000} attenuation curve, modified to allow its slope to vary and to add a UV bump (see section 3.4 of \citealt{SALIM+2018}).

The SED estimation was additionally constrained by the galaxies' total IR luminosities (i.e. matching the energy absorbed by the dust within galaxies with the energy it re-emits; see section 3.2 of \citealt{SALIM+2018}). Total IR luminosities were derived from the $22$ $\mu$m WISE photometry (if available, $12$ $\mu$m if not) using \cite{CHARY+2001} templates, further corrected based on Herschel \citep{VALIANTE+2016} IR photometry (see section 3.1 of \citealt{SALIM+2018}). The overall quality of fit was measured by its reduced chi-squared value ($\chi^{2}_{r}$).

Astrophysical features including rest-frame absolute magnitudes, colour excesses [$E(B-V)$], stellar masses ($M_{*}$), stellar metallicities ($Z$), mass-weighted stellar ages ($MWSA$), and specific star formation rates [$sSFR$ (SED)] were derived from the full ensemble of possible synthetic spectra via a Bayesian approach \citep{SALIM+2007}. The likelihood of the fit of each synthetic spectrum to the photometry of each galaxy was used to generate a probability density function for each feature, with the likelihood-weighted means of the functions being quoted as the best estimates of the features, and the likelihood-weighted standard deviations as the errors. 

\subsubsection{Final low-redshift sample}
\label{subsubsec:finalg}

Our final GSWLC-2 sample is subject to the following selections. Firstly, we only retain galaxies whose best-fitting \texttt{CIGALE} SEDs produce $\chi^2_{r} <= 11.07$ (i.e. the mean plus two standard deviations of the logarithmic GSWLC-2 distribution in $\chi^2_{r}$), in order to omit particularly poorly constrained fits. Spectroscopic redshifts are limited to the range $0.02 < z < 0.08$, and stellar masses (as estimated via Bayesian analysis of the synthetic \texttt{CIGALE} spectra) to $> 10^{9.5}$ M$_{\odot}$. These two restrictions ensure completeness above the imposed stellar mass limit. Finally, broad-line active galactic nuclei are removed by asserting \texttt{flag\_sed} $=0$. Our final GSWLC-2 sample has a median redshift of $0.06$ and contains $177,362$ galaxies.

As additional, \texttt{CIGALE}-independent indicators of the stellar populations in GSWLC-2 galaxies, we invoke \cite{BRINCHMANN+2004} specific star formation rates [$sSFR$ (ind.)] and $4000$ \r A break strengths [$D(4000)$]. The SFRs sum two components: a spectroscopic fibre SFR, and a photometric SFR outside the fibre, given by an optical SED fit \citep{SALIM+2007}. The fibre SFR is given by either a H$\alpha$ calibration \citep{CHARLOT+2001} or, in the case of spectra that have a contribution from an active galactic nucleus, a $D(4000)$-based estimate (itself calibrated on the emission lines of pure star-forming galaxies). These SFRs are then normalised by photometrically-determined stellar masses to give $sSFR$ (ind.). The timescale probed by $sSFR$ (ind.) lies between the $10$ Myr timescale of the H$\alpha$-calibrated fibre SFRs, and the $1$ Gyr timescale of optical SED-based SFRs \citep{SALIM+2016}. The $D(4000)$ measurements apply to fibre region only. Both of these features are available for $97$ per cent of the galaxies in our GSWLC-2 sample.

We obtain S\'ersic indices ($n_{g}$) and circularised half-light radii ($R_{1/2}$) for the galaxies in our GSWLC-2 sample from catalogues assembled by \cite{SIMARD+2011}. Both were derived from fits of singular \cite{SERSIC-1963,SERSIC-1968} profiles to $r$-band images of galaxies in SDSS. The S\'ersic indices have minimum and maximum allowed values of $0.5$ and $8$ respectively. S\'ersic indices and half-light radii are available for $96.2$ per cent of the galaxies in our final GSWLC-2 sample. We also use \cite{SIMARD+2011} $r$-band bulge-to-total ratios ($B/T_{r}$) for these galaxies, which were based on fits consisting of two components: a S\'ersic bulge (fixed at an index of $4$) and an exponential disc. Local environmental densities, available for $92.1$ per cent of our GSWLC-2 galaxies, come from \cite{BALDRY+2006}. They averaged the surface densities of SDSS galaxies with respect to their fourth- and fifth-nearest density-defining neighbour within $1,000$ km s$^{-1}$ along the line of sight. We calculate local overdensities ($\delta$) using $\delta=(\Sigma-\bar{\Sigma})/\bar{\Sigma}$, where $\Sigma$ is the local surface density and $\bar{\Sigma}$ the average surface density of the sample.

\subsection{VIPERS}
\label{subsec:vipers}

The VIMOS Public Extragalactic Redshift Survey (VIPERS; \citealt{GUZZO+2014,GARILLI+2014,SCODEGGIO+2018}) aimed to match the statistical fidelity of low-redshift surveys like SDSS, but at intermediate redshifts ($z \sim 0.7$). The survey was conducted using the VIMOS spectrograph \citep{LEFEVRE+2003} of the European Southern Observatory's Very Large Telescope. Its targeting was based on the Canada-France-Hawaii Telescope Legacy Survey Wide (CFHTLS-Wide) photometric catalogue\footnote{\url{http://www.cfht.hawaii.edu/Science/CFHLS}}, with objects qualifying for VIPERS if they had extinction-corrected $i$-band magnitudes $i_{AB} < 22.5$. An additional $ugri$ colour cut was applied to remove low-redshift ($z \lesssim 0.5$) galaxies from the survey \citep{GUZZO+2014}. PDR2, the second and final public data release of VIPERS, comprises spectroscopy for $97,414$ objects \citep{SCODEGGIO+2018}. $52,114$ of these objects ($51,522$ galaxies and $592$ broad-line active galactic nuclei) have `secure' ($>99$ \% confidence) redshifts. This secure-redshift sample was the subject of the \cite{SIUDEK+2018a} study, and is the basis of our present VIPERS sample\footnote{The use of these secure redshifts is recommended by \cite{GARILLI+2014} and \cite{SCODEGGIO+2018} for scientific analyses. Approximately $75$ per cent of all VIPERS galaxies within and throughout the redshift range of our final VIPERS sample (see Section \ref{subsubsec:finalv}) have secure redshifts (see figure 9 of \citealt{SCODEGGIO+2018}).}.

Photometry for this sample was taken from a catalogue prepared by \cite{MOUTARD+2016a}. The CFHTLS-Wide photometric catalogue (i.e. the basis of the targeting for VIPERS) provided optical photometry for this sample in $u*$, $g$, $r$, $i$, and $z$ bands. \cite{MOUTARD+2016a} derived total magnitudes for the galaxies in this sample by rescaling their isophotal magnitudes. These isophotal magnitudes were chosen for the accuracy of their colours with a view to photometric redshift estimation; this choice now benefits our SED estimation as well.

Like for our GSWLC-2 sample, UV photometry came from GALEX. \cite{MOUTARD+2016a} supplemented existing Deep Imaging Survey observations of VIPERS galaxies with deep GALEX observations of their own in order to improve UV coverage within the VIPERS footprint. Coverage is complete in the W1 field of VIPERS, but not in the W4 field (see figure 1 of \citealt{MOUTARD+2016a}). UV photometry was then measured using a Bayesian approach with the $u*$-band profiles of galaxies as priors \citep{CONSEIL+2011}, which mitigated the confusion of sources due to their blended UV profiles. UV photometry in at least one of GALEX's two bands (almost always $NUV$ if just one) is available for $52$ per cent of galaxies in the \cite{SIUDEK+2018a} sample and in our final VIPERS sample (Section \ref{subsubsec:finalv}).

Near-infrared (NIR) $K_{s}$-band photometry came from a dedicated CFHT WIRCam \citep{PUGET+2004} follow-up survey of VIPERS galaxies \citep{MOUTARD+2016a}. This $K_{s}$-band photometry was validated against NIR photometry from the VISTA Deep Extragalactic Observations (VIDEO) survey \citep{JARVIS+2013}, exhibiting good agreement. We also take VIDEO survey $Z$, $Y$, $J$, $H$, and $K_{s}$ NIR photometry for our sample where available ($11$ per cent of the \citealt{SIUDEK+2018a} sample, $10$ per cent of our final VIPERS sample; Section \ref{subsubsec:finalv}). CFHT $K_{s}$-band photometry is available for $91$ per cent of galaxies in the \cite{SIUDEK+2018a} sample, and for $93$ per cent of galaxies in our final VIPERS sample (Section \ref{subsubsec:finalv}).

\subsubsection{VIPERS rest-frame SEDs}
\label{subsubsec:vseds}

The SEDs of VIPERS galaxies are estimated via a full fit of synthetic \texttt{CIGALE} spectra to the available UV-through-NIR photometry. This differs slightly from the method used for the GSWLC-2, whose NIR SEDs were constrained not by their shapes but simply by their total IR luminosities (Section \ref{subsubsec:gseds}). While we use the same stellar templates (\citealt{BRUZUAL+2003}, with \citealt{CHABRIER-2003} initial mass functions and metallicities of $0.004$, $0.008$, $0.02$, or $0.05$) for VIPERS as were used for GSWLC-2 , the SFHs are adjusted to reflect the change in cosmic epoch between samples and to account for the possibility of very recent bursts of star formation\footnote{Consequences of this adjustment are discussed in Section \ref{subsubsec:pass}; the properties of most VIPERS galaxies appear accurate, except for those a subpopulation of passive VIPERS galaxies.}. Astrophysical features are derived for VIPERS galaxies using the same Bayesian approach as for GSWLC-2 galaxies (see Section \ref{subsubsec:gseds}).

\subsubsection{Final intermediate-redshift sample}
\label{subsubsec:finalv}

We make the following selections to yield our final VIPERS sample. Galaxies are kept if the $\chi^{2}_{r}$ of their best-fitting \texttt{CIGALE} SED has a value less than or equal to the mean plus two standard deviations ($=18.85$) of the overall logarithmic VIPERS distribution. Spectroscopic redshifts are restricted to being within the range $0.5 < z < 0.8$, balancing our intent to define a co-eval population of galaxies against the need to keep the sample as large as possible. Like our GSWLC-2 sample, stellar masses are limited to $> 10^{9.5}$ M$_{\odot}$ with a view to mass completeness (though see Sections \ref{subsubsec:pass} and \ref{subsec:env}, where we discuss shortcomings). Broad-line active galactic nuclei and serendipitous secondary spectral sources are removed using \texttt{zflag} $< 10$. Ultimately, this gives us a final VIPERS sample consisting of $31,889$ galaxies, with a median redshift of $0.65$. 

Emission-line SFRs, which are independent of our \texttt{CIGALE} SED estimation, were calculated from the [OII] $\lambda 3727$ fluxes of the galaxies in our VIPERS sample using the calibration (which includes empirical stellar-mass-based corrections) of \cite{GILBANK+2010,GILBANK+2010e,GILBANK+2011}. These [OII] $\lambda 3727$ fluxes are available for $27,537$ of the galaxies in our VIPERS sample, and they probe short timescales of star formation ($\sim 10$ Myr). We normalise these [OII] SFRs by our \texttt{CIGALE} stellar masses to yield specific star formation rates\footnote{Our use of stellar masses given by \texttt{CIGALE} means that these $sSFR$ (ind.) estimates are not entirely independent of \texttt{CIGALE}, however we expect that \texttt{CIGALE}'s stellar masses would be consistent with those estimated via other methods, given that stellar mass estimates are generally quite robust \citep{BELL+2001}.} [$sSFR$ (ind.)]. $D(4000)$ was measured from VIPERS spectra by \cite{GARILLI+2014}, using the same \cite{BALOGH+1999} method as was used for SDSS \citep{BRINCHMANN+2004}. S\'ersic indices and circularised half-light radii for the galaxies in our VIPERS sample are given by \cite{KRYWULT+2017}, who fitted the $i$-band light distributions of galaxies with single \cite{SERSIC-1963,SERSIC-1968} profiles. These features are available for $96.2$ per cent of the galaxies in our final VIPERS sample. We winsorise the S\'ersic indices to values of $0.5$ and $8$ in order to match our GSWLC-2 sample. The overdensities of $91.7$ per cent VIPERS galaxies were derived by \cite{CUCCIATI+2017}, based on fifth-nearest neighbour surface densities.

\section{Clustering method}
\label{sec:meth}

We apply the Fisher Expectation-Maximisation algorithm, which estimates the parameters of the Discriminative Latent Mixture model. \cite{BOUVEYRON+2012} offer full, rigorous, mathematical derivations of both the Discriminative Latent Mixture model and the Fisher Expectation-Maximisation algorithm in their paper; here, we offer brief summaries of the model (Section \ref{subsec:dlm}), and of its implementation via the algorithm (Section \ref{subsec:fem}). In Section \ref{subsec:mdl}, we discuss some additional relevant practicalities to the use of the model and algorithm, and in Section \ref{subsec:feats}, we describe the shared feature space within which we cluster our two samples.

\subsection{The Discriminative Latent Mixture model}
\label{subsec:dlm}

The Discriminative Latent Mixture (DLM) model is a clustering approach that incorporates dimensionality reduction on the fly to determine a frugal fit to the structure of an input sample, which is assumed to consist of $k$ clusters. Selection of the value of $k$ is discussed in Section \ref{subsec:mdl}.

The key premise of the DLM model is thus: a sample represented in a $D$-dimensional space that is defined by observed features actually occupies an intrinsic $d$-dimensional subspace ($d < D$; the `empty space phenomenon'; \citealt{SCOTT+1983}) that is defined by unobserved, latent features. Hence, the clustering structure of the sample should be fitted in this intrinsic subspace.

\begin{figure}
\centering
\includegraphics[width=0.45\textwidth]{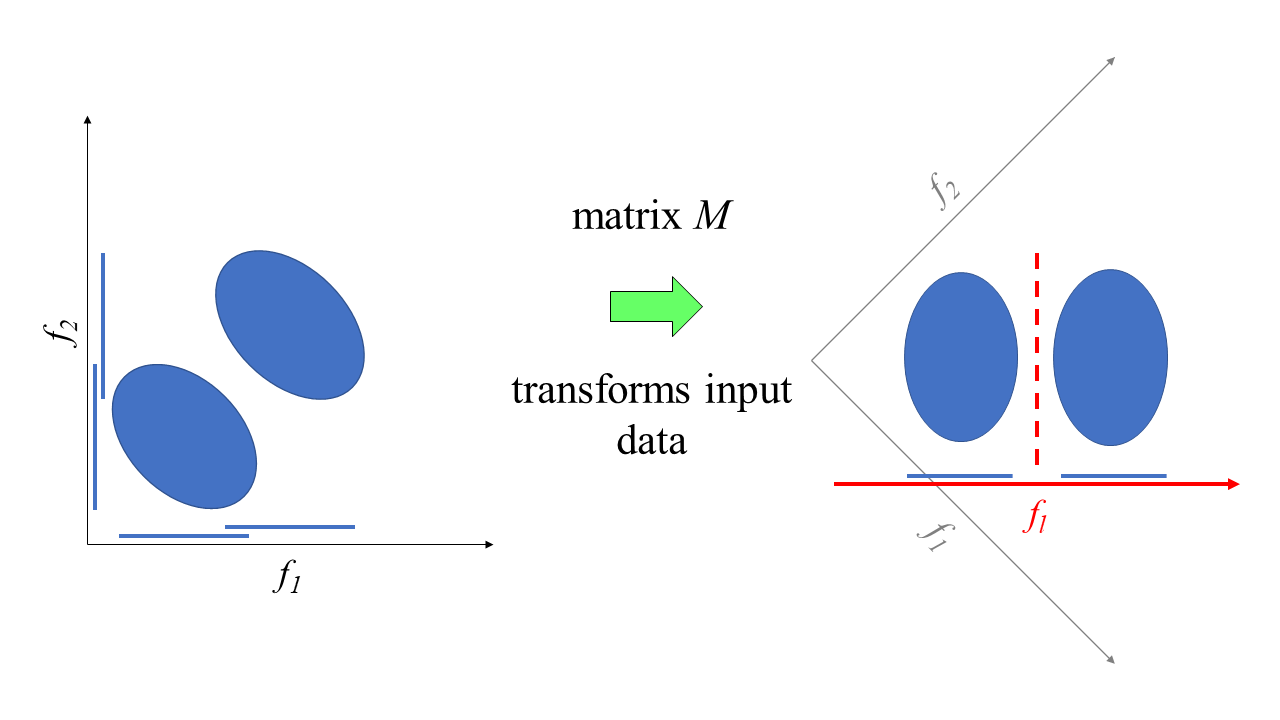}
\caption{A simple demonstration of the principles behind subspace clustering. Here, a sample consisting of two clusters (represented by the two blue ellipses) is represented in a two-dimensional full space defined by features $f_{1}$ and $f_{2}$. Matrix $M$ enables the transformation of the sample to a one-dimensional subspace, defined by latent feature $f_{l}$, in which the two clusters are easily discriminated.}
\label{fig:dlm}
\end{figure}

The subspace has two important properties in the context of the DLM model. Firstly, of all possible $d$-dimensional subspaces, it is the one that best discriminates the $k$ clusters in the sample. The model assumes $1 \leq d \leq k-1$: that $k$ clusters may be distinguished in $k-1$ dimensions or fewer (see Section \ref{subsec:mdl} for further explanation). Secondly, the subspace is linearly related to the full $D$-dimensional space, such that the unobserved, latent features are linear combinations of the observed features. Hence there exists a matrix $M$, common to all of the $k$ clusters, that enables the transformation of the sample between the full space and the subspace. This transformation matrix is constrained by the condition that the basis vectors of the subspace must be orthonormal. Estimation of the transformation matrix $M$ is explained in Section \ref{subsec:fem}. Selection of the value of $d$ is explained in Section \ref{subsec:mdl}. Fig. \ref{fig:dlm} demonstrates these two important properties of the subspace.

The DLM model assumes that the sample is distributed among a mixture of $k$ Gaussian density functions within the discriminative latent subspace. The functions, each of which corresponds to a cluster, are defined by three parameters: a mean vector ($\mu_{k}$), a covariance matrix ($\Sigma_{k}$), and a scalar relative mixture proportion ($\pi_{k}$). The matrix $M$ enables the transformation of these parameters back to the full space. For the covariances, this includes the addition of Gaussian `noise' ($\delta_{k}$; unique to each of the clusters), which is defined as non-discriminative structure that exists in the full space but not in the subspace. While $\Sigma_{k}$ captures the cluster covariances inside the discriminative latent subspace, $\delta_{k}$ captures the cluster covariances outside the subspace. Full space covariances are the sum of both. Estimation of the cluster means, covariances, and noise terms is discussed in Section \ref{subsec:fem}.

Implementation of the DLM model hence requires the estimation of the following parameters: 

\begin{itemize}
    \item $k-1$ relative mixture proportions ($\pi_{k}$; given that one cluster has a proportion of $1$);
    \item $kd$ parameters for the mean vectors ($\mu_{k}$) in the subspace;
    \item $kd(d+1)/2$ parameters for the covariance matrices ($\Sigma_{k}$) in the subspace (fewer than $kd^{2}$ parameters because covariance matrices are symmetric);
    \item $d(D-(d+1)/2)$ parameters for the transformation matrix $M$ (the number of free parameters, given the constraint that the basis vectors of the subspace must be orthonormal);
    \item $k$ noise terms ($\delta_{k}$; given that this non-discriminative structure is Gaussian and spherical, and may therefore by parametrised by a single value in reference to the Gaussian density function estimated for each cluster).
\end{itemize}

The total number of parameters ($q_{DLM}$) is most strongly influenced by the value of $d$. The maximum $q_{DLM}$ at a certain combination of $D$ and $k$ is given by setting $d$ to its maximum value of $k-1$ (based on the aforementioned assumption that $k$ clusters may be distinguished in $k-1$ dimensions or fewer). $q_{DLM}$ is smaller than the number of parameters that must be estimated for a Gaussian Mixture Model in the full space ($q_{GMM}$), especially if $d << D$ ($q_{GMM}$ is given by the sum of $k-1$ relative mixture proportions, $kD$ parameters for the mean vectors, and $kD(D+1)/2$ parameters for the covariance matrices).

Parameter $q_{DLM}$ may be further reduced by imposing additional constraints upon the DLM model. For example, the covariance matrices ($\Sigma_{k}$) may be assumed to be the same for all Gaussians ($\Sigma$; the Gaussians all have the same shape). Alternatively, they may be assumed to be diagonal ($\alpha_{k,j}$, where the subscript $j$ indicates a different variance in each dimension of the subspace), meaning the latent features that define the subspace are uncorrelated. These diagonal covariance matrices may then also be assumed to be isotropic ($\alpha_{k}$; spherical Gaussians in the subspace), the same for all Gaussians ($\alpha_{j}$), or both ($\alpha$). The noise terms ($\delta_{k}$) may be assumed to be the same for all Gaussians ($\delta$) as well. Constraints like these may be imposed to speed up the clustering, in anticipation of a particular clustering structure, or (as in our case) to compare fits of models of varying complexities (see also Section \ref{subsec:mdl}). The various combinations of these constraints on the covariance matrices and noise terms yield $11$ submodels of the full $\Sigma_{k}$, $\delta_{k}$ DLM model. They are listed in full in table 1 of \cite{BOUVEYRON+2012} (and listed partially in Table \ref{tab:icl} of this paper).

\subsection{The Fisher Expectation-Maximisation algorithm}
\label{subsec:fem}

The Fisher Expectation-Maximisation algorithm (\texttt{Fisher-EM} or, as we will call it in this paper, \texttt{FEM}) estimates the parameters ($\pi_{k}$, $\mu_{k}$, $\Sigma_{k}$, $M$, $\delta_{k}$) of the DLM model, fitting a sample of $N$ observations, observed in a $D$-dimensional space (the `full' space, defined by $D$ observed features), with $k$ Gaussian density functions in a $d$-dimensional discriminative latent subspace ($1 \leq d \leq k - 1$). \texttt{FEM} comprises the following steps:

\begin{enumerate}
\setcounter{enumi}{-1}
\item Initialisation: $k$ starting points are selected within the extent of the sample in the full space;
\item Expectation (E): transform the parameters of the mixture of Gaussians to the full space, and calculate the probability of each observation having originated from each Gaussian;
\item Fisher (F; based on discriminant analysis): using the observation probabilities, find the subspace that best separates the Gaussians;
\item Maximisation (M): update the parameters of the mixture of Gaussians (including non-discriminative structure, termed `noise') within the subspace.
\end{enumerate}

The Expectation, Fisher, and Maximisation steps are iterated such that \texttt{FEM} improves its estimates of the DLM model parameters as it proceeds. \texttt{FEM} is slow to run on our large samples and, unlike traditional expectation-maximisation algorithms, does not always converge perfectly (such that there are no changes between successive iterations; due to the Fisher step). We therefore terminate \texttt{FEM} at the completion of $25$ iterations; changes between iterations become negligible well before this number (see Appendix \ref{app:iter}). The final output of \texttt{FEM} is a series of $k$ probabilities for each of the observations: probabilities of each observation having originated from each of the $k$ Gaussians. Final cluster labels are given by assigning each observation to the Gaussian with the highest probability of having originated it.

While successive iterations of \texttt{FEM} improve its estimates of the DLM model parameters, these estimates improve only towards local maxima of the likelihood function. \texttt{FEM} is hence run with varying initialisations, which may intuitively be considered as `exploring the surface' of the likelihood function of the model parameters. This encourages optimisation towards different local maxima and, hopefully among these, the global maximum, corresponding to the very best estimate of the DLM model parameters.

Initialisation techniques may be as simple as a uniform random selection of $k$ observations from the sample. We opt to use the \texttt{k-means} algorithm \citep{MACQUEEN-1967,LLOYD-1982}, which implements a simple centroid-based clustering approach, to generate initialisations for \texttt{FEM}. \texttt{k-means} is an expectation-maximixation algorithm and, like \texttt{FEM}, only optimises to local maxima. We therefore initialise \texttt{k-means} \textit{itself} $100$ times in the hope of encouraging optimisation towards the global maximum of \textit{its} objective function (which measures how separated the clusters are). Use of varying initialisations provided by a heuristic like \texttt{k-means} leads to `pre-optimisation' of \texttt{FEM} because the separated centroids are likely to span the full extent of the sample in its full space. This facilitates improvement of \texttt{FEM}'s estimates of the DLM model parameters towards the global maximum of their likelihood functions. Following this initialisation, \texttt{FEM} proceeds to the Fisher step, in which it finds the subspace that best separates the final \texttt{k-means} clusters, and to the Maximisation step, in which it fits the observations with a mixture of Gaussians within this subspace. \texttt{FEM} then loops back around to the Expectaton step and begins iterating proper.

The Expectation step uses the parameters estimated in the Maximisation step ($\pi_{k}$, $\mu_{k}$, $\Sigma_{k}$, $\delta_{k}$) to calculate the conditional probability of each observation having originated from each of the $k$ Gaussians. These parameters are transformed from the subspace, within which they are estimated in the Maximisation step, to the full space using matrix $M$, found in the Fisher step.

The Fisher step finds the $d$-dimensional discriminative latent subspace that best separates the new partition calculated in the Expectation step. \cite{BOUVEYRON+2012} base this step on discriminant analysis, which finds the linear combination of the input features that maximises the ratio of the scatter \textit{between} clusters to the scatter \textit{within} clusters. Similar principles have been applied for the visualisation of multi-dimensional clusters as well (e.g. \citealt{LISBOA+2008}). These scatters are weighted by the probabilities calculated in the Expectation step. A constraint of the DLM model is that the $d$ basis vectors that define the subspace must be orthonormal, which is not necessarily a property of the $d$ basis vectors that linear discriminant analysis (LDA) provides. \cite{BOUVEYRON+2012} assert this constraint by applying the orthonormal discriminant vector method (ODV; \citealt{OKADA+1985}). ODV uses LDA to find the $d$ basis vectors in succession while also ensuring the orthonormality of each new basis vector with respect to all of those that have already been calculated. The first basis vector, which is free of this constraint, is given by the direct application of LDA to the sample in the full space. The $d$ orthonormal basis vectors constitute the columns of $M$, the matrix that enables the transformation of the sample between the full space and the subspace.

The Maximisation step updates the estimates of the means, covariances, and relative mixture proportions ($\pi_{k}$, $\mu_{k}$, $\Sigma_{k}$) of the $k$ Gaussians in order to maximise the likelihood of the fit. These estimates are measured within the subspace found in the Fisher step, and are weighted by the probabilities calculated in the Expectation step. This step also updates the estimates of the noise terms ($\delta_{k}$), which is given by the differences between the full-space variances (again weighted by the probabilities calculated in the Expectation step) and the newly updated subspace variances.

\subsection{Practicalities}
\label{subsec:mdl}

We do not presume a DLM submodel or value of $k$ with which to fit our samples. Instead, we conduct a search over all of the DLM submodels and over a range of values of $k$ to determine the best-fitting combination. Three of the DLM submodels ($\alpha_{j}$, $\delta_{k}$; $\alpha_{j}$, $\delta$; $\alpha$, $\delta_{k}$) are not available for use in the version of \texttt{FEM}\footnote{Version 1.5.1, for the \texttt{R} statistical computing environment.} that we use for our fitting. This reduces the total number of available submodels from $12$ (including the full $\Sigma_{k}$, $\delta_{k}$ model) to nine.

We identify the best-fitting combination of DLM submodel and value of $k$ by using the Integrated Completed Likelihood criterion (ICL; \citealt{BIERNACKI+2000}):

\begin{equation}
\mathrm{ICL} = \ln(L) - \frac{q_{DLM}}{2} \ln(N) - [ - \Sigma_{i=1}^{N}\ \Sigma_{l=1}^{k}\ z_{i,l} \ln(p_{i,l})],
\label{eq:icl}
\end{equation}

where $L$ is the likelihood of the fit, $p_{i,l}$ is the probability of observation $i$ belonging to cluster $l$, and $z_{i,l}$ denotes cluster membership, taking a value of $1$ when $p_{i,l} = \mathrm{max}(p_{i,:})$ and a value of $0$ otherwise. The ICL is related to the popular Bayesian Information criterion (BIC; \citealt{SCHWARZ-1978}). While both the BIC and ICL criteria penalise the likelihood using the number of model parameters (to avoid over-fitting), the ICL criterion also rewards separated clusters (a general aim of clustering). The combination of submodel and $k$ that returns the highest ICL score is deemed the the best fit.

The dimensionality of the discriminative latent subspace is constrained by the number of clusters being fitted: $1 \leq d \leq k-1$. The maximal $d = k - 1$ case may intuitively be understood as setting the origin of the subspace at one of the $k$ cluster centres so that the full-space vectors to each of the remaining $k - 1$ cluster centres define the basis vectors of the subspace. If multiple clusters lie along the same direction in the full space, the number of basis vectors needed to define the subspace is reduced. In our application of \texttt{FEM}, we hold $d$ at its maximum value of $k-1$. This is recommended by \cite{BOUVEYRON+2012} to avoid omitting any discriminative structure from the subspace and to ease convergence of \texttt{FEM} (which may become unstable or fail to converge if $d$ is too small in comparison with $k$ and/or $D$). Hence, the maximum value of $k$ in our model selection search is $9$ (set by $d=8$, given $D=9$).

\subsection{Input features to the clustering}
\label{subsec:feats}

The fitting of the clustering structures of both of our samples is conducted within a nine-dimensional feature space defined by UV-through-NIR colours. We opt for colours because of their widespread use in studies of galaxy evolution, and because of the relative ease with which they may be measured. While clustering in terms of derived astrophysical features may facilitate a more direct interpretation of resultant clusters in terms of theories of galaxy evolution, their derivation is much more model-dependent than that of colours. Clustering in terms of colours ensures the generalisability of our outcomes.

The colours that we use are calculated not from the observed photometry that is used as input to the SED fitting, but from rest-frame magnitudes estimated by \texttt{CIGALE}. This ensures homogeneity among the input features, and that the feature space is defined by rest-frame colours (which is more difficult to ensure using colours calculated directly from observed photometry). In addition, the SED estimation can infer the rest-frame magnitudes of galaxies in bands for which there is no observed photometry. The full list of rest-frame colours used for the clustering is: $FUV-NUV$, $NUV-u$, $u-g$, $g-i$, $i-r$, $r-z$, $z-J$, $J-H$, and $H-K_{s}$. These rest-frame colours are intended to represent the shape of each galaxy's UV-through-NIR SED, and to remove the influence of the intrinsic brightnesses of the galaxies on the clustering outcomes. The rest-frame magnitudes of GSWLC-2 galaxies (but not VIPERS galaxies) are subject to some smoothing (see Appendix \ref{app:sm}). In addition, the rest-frame NIR colours of GSWLC-2 galaxies were inferred from UV and optical photometry (given the lack of input NIR photometry). Use of the term `colour' from this point forward in this paper is intended in reference to these rest-frame colours, as estimated by \texttt{CIGALE}.

These colours differ from those used by \cite{SIUDEK+2018a}; they used rest-frame colours defined with reference to the rest-frame $i$-band magnitudes of galaxies ($FUV - i$, $NUV - i$, etc.), also with the aim of removing the influence of galaxy intrinsic brightnesses on their clustering outcomes. However, their UV colours, defined across the largest distances in wavelength among their features, exhibited large spreads (up to a factor of $10$ larger than the spreads of other colours) and dictated much of their clustering. Preliminary tests of clustering with these $i$-band based colours for our present, carefully prepared samples confirmed this. The $\alpha_{k,j}$ and $\alpha_{k,j}$ submodels achieved the highest ICL scores for these $i$-band colours, but gave only relatively crude segmentations of our samples (see also Appendix \ref{app:icl}). Our colours, defined using magnitudes in filters at neighbouring effective wavelengths, mitigate this effect and encourage \texttt{FEM} to converge to more detailed partitions (although, as shown in Fig. \ref{fig:imp}, bluer colours are still most important).

\begin{table*}
\caption{Integrated Completed Likelihood (ICL) scores reported by our search over all possible combinations of submodel (see Section \ref{subsec:dlm} for further explanation) and $k$ for our samples. The uncertainties span the full range of ICL scores (i.e. from minimum to maximum) registered over $100$ initialisations for each combination. As mentioned in Section \ref{subsec:mdl}, only nine of the $12$ submodels are available in the version of \texttt{FEM} that we use for our fitting. The score of the best-fitting combination is highlighted using bold text. While submodel $\Sigma$, $\delta$ produces the highest score for our GSWLC-2 sample (at $k=9$), we reject it for reasons given in Appendix \ref{app:icl}. Blank entries correspond to combinations for which \texttt{FEM} did not converge (see Appendix \ref{app:icl}). The entries listed in this table are subject to the multipliers at the right-hand side of each section. The ICL scores for our GSWLC-2 sample are systematically higher than those for our VIPERS sample because it contains more galaxies.}
\label{tab:icl}
\centering
\begin{tabular}{c c | r r r r r r r r r | c }
\hline
                                       & & \multicolumn{9}{| c |}{Submodel} & \\
\cline{3-11}
                                       & & $\Sigma_{k}$, $\delta_{k}$ & $\Sigma_{k}$, $\delta$ & $\Sigma$, $\delta_{k}$ & $\Sigma$, $\delta$ & $\alpha_{k,j}$, $\delta_{k}$ & $\alpha_{k,j}$, $\delta$ & $\alpha_{k}$, $\delta_{k}$ & $\alpha_{k}$, $\delta$ & $\alpha$, $\delta$ &\\
\hline
\multirow{8}{*}{\rotatebox{90}{GSWLC-2}} & $k=2$ & 1.8 $\pm$ 0.0 & 1.5 $\pm$ 0.0 & 0.4 $\pm$ 0.0 & -6.2 $\pm$ 0.0 & 1.8 $\pm$ 0.0 & -4.7 $\pm$ 0.0 & 1.8 $\pm$ 0.0 & -4.7 $\pm$ 0.0 & -6.2 $\pm$ 0.0 & \multirow{8}{*}{$\times 10^{5}$}\\
                                         & $k=3$ & 8.2 $\pm$ 0.0 & 7.8 $\pm$ 0.0 & -141.0 $\pm$ 0.0 & 0.0 $\pm$ 0.0 & 3.8 $\pm$ 0.0 & -5.1 $\pm$ 0.0 & 3.7 $\pm$ 0.0 & -5.3 $\pm$ 0.0 & -4.5 $\pm$ 0.0 & \\
                                         & $k=4$ & 11.7 $\pm$ 0.0 & 11.3 $\pm$ 0.0 & 2.7 $\pm$ 0.0 & 5.2 $\pm$ 0.0 & & & 4.9 $\pm$ 0.0 & -4.2 $\pm$ 0.0 & -3.8 $\pm$ 0.0 & \\
                                         & $k=5$ & 13.4 $\pm$ 1.4 & 13.4 $\pm$ 0.2 & -46.2 $\pm$ 51.5 & 8.7 $\pm$ 0.4 & 6.7 $\pm$ 1.0 & & 6.0 $\pm$ 0.0 & -2.2 $\pm$ 0.0 & -5.7 $\pm$ 1.3 & \\
                                         & $k=6$ & 16.7 $\pm$ 0.0 & & 9.4 $\pm$ 0.0 & 13.0 $\pm$ 0.1 & & & 6.8 $\pm$ 0.0 & 0.7 $\pm$ 0.0 & -5.2 $\pm$ 0.0 & \\
                                         & $k=7$ & \textbf{17.9 $\pm$ 0.2} & & 11.8 $\pm$ 2.0 & 14.2 $\pm$ 1.6 & & & 7.2 $\pm$ 0.0 & 2.3 $\pm$ 0.0 & -5.6 $\pm$ 0.0 & \\
                                         & $k=8$ & & & & 16.3 $\pm$ 1.3 & & & 8.1 $\pm$ 0.0 & 1.1 $\pm$ 0.0 & -5.7 $\pm$ 0.0 & \\
                                         & $k=9$ & & & 17.1 $\pm$ 0.0 & \sout{\textbf{18.1 $\pm$ 0.9}} & & & 8.0 $\pm$ 0.0 & 3.9 $\pm$ 0.0 & -6.0 $\pm$ 0.0 & \\
\hline
\multirow{8}{*}{\rotatebox{90}{VIPERS}}  & $k=2$ & & 2.1 $\pm$ 0.0 & 4.3 $\pm$ 0.0 & -8.4 $\pm$ 0.0 & & -8.0 $\pm$ 0.0 & & -8.0 $\pm$ 0.0 & -8.4 $\pm$ 0.0 & \multirow{8}{*}{$\times 10^{4}$}\\
                                         & $k=3$ & & 11.1 $\pm$ 0.0 & -421.0 $\pm$ 0.0 & -0.4 $\pm$ 0.0 & & -4.7 $\pm$ 0.0 & 6.8 $\pm$ 0.0 & -5.3 $\pm$ 0.0 & -7.9 $\pm$ 0.0 &\\
                                         & $k=4$ & & & -294.0 $\pm$ 0.0 & 8.3 $\pm$ 0.0  & & -6.7 $\pm$ 0.0 & 8.3 $\pm$ 0.0 & -7.4 $\pm$ 0.0 & -9.0 $\pm$ 0.0 &\\
                                         & $k=5$ & 32.9 $\pm$ 0.0 & 32.4 $\pm$ 0.0 & 15.6 $\pm$ 0.0 & 16.3 $\pm$ 0.2 & & & 10.5 $\pm$ 0.0 & -2.9 $\pm$ 0.0 & -7.0 $\pm$ 0.0 &\\
                                         & $k=6$ & & & 20.9 $\pm$ 1.5 & 23.6 $\pm$ 0.0 & & & 13.0 $\pm$ 0.0 & 2.9 $\pm$ 0.0 & -3.8 $\pm$ 0.1 &\\
                                         & $k=7$ & & \textbf{41.8 $\pm$ 0.1} & 26.4 $\pm$ 2.3 & & & & 15.9 $\pm$ 0.8 & 7.1 $\pm$ 0.4 & -6.7 $\pm$ 0.0 &\\
                                         & $k=8$ & & & & & & & 14.6 $\pm$ 0.0 & 3.2 $\pm$ 0.0 & -10.8 $\pm$ 0.0 &\\
                                         & $k=9$ & & & & & & & 12.5 $\pm$ 0.0 & & -10.8 $\pm$ 0.0 &\\
\hline
\end{tabular}
\end{table*}

\section{Results}
\label{sec:res}

\subsection{\texttt{FEM} submodel selection}
\label{subsec:best}

As outlined in Section \ref{subsec:mdl}, we conduct a search for the best-fitting \texttt{FEM} submodel and number of clusters for our samples. We identify the best-fitting combination using the ICL criterion (Equation \ref{eq:icl}), which penalises the number of parameters of the submodel while favouring separated clusters. Table \ref{tab:icl} lists ICL scores reported for both samples. The uncertainties on these scores, which span the \textit{full} variation (i.e. from minimum to maximum) over $100$ initialisations, show that \texttt{FEM} is extremely stable and self-consistent, robustly converging to highly similar outcomes over successive runs that use the same combination of submodel and number of clusters. The best-fitting combinations for each sample are highlighted using bold text. We briefly describe patterns of behaviour of the various submodels and explain the large range in ICL scores in Appendix \ref{app:icl}. Despite it registering the highest score for the GSWLC-2 sample, we reject the $k=9$, $\Sigma$, $\delta$ combination due to its inclusion of empty clusters (explained further also in Appendix \ref{app:icl}).

Both samples are best partitioned into seven clusters, within a six-dimensional discriminative latent subspace. The Gaussian density functions representing the clusters are each described by their own unique, full covariance matrices ($\Sigma_{k}$); the clusters each have different shapes, and the use of full covariance matrices indicates correlations (as expected) among the input features within the subspaces. While the best-fitting submodel for the GSWLC-2 sample uses unique noise terms for each cluster ($\delta_{k}$), the best-fitting submodel for the VIPERS sample does not ($\delta$), owing to the smoother distribution of the VIPERS sample in the feature space (see e.g. Fig. \ref{fig:sub}). Submodels $\Sigma_{k}$, $\delta_{k}$ and $\Sigma_{k}$, $\delta$ report similar ICL scores and produce similar clustering structures in general and may therefore readily be compared with one another (see also Appendix \ref{app:icl}). That \texttt{FEM} has converged to highlighting these closely related submodels as being optimal for describing both samples is encouraging, and gives us confidence that we are conducting a fair comparison.

\subsection{Feature importance}
\label{subsec:fi}

\begin{figure}
\centering
\includegraphics[width=0.45\textwidth]{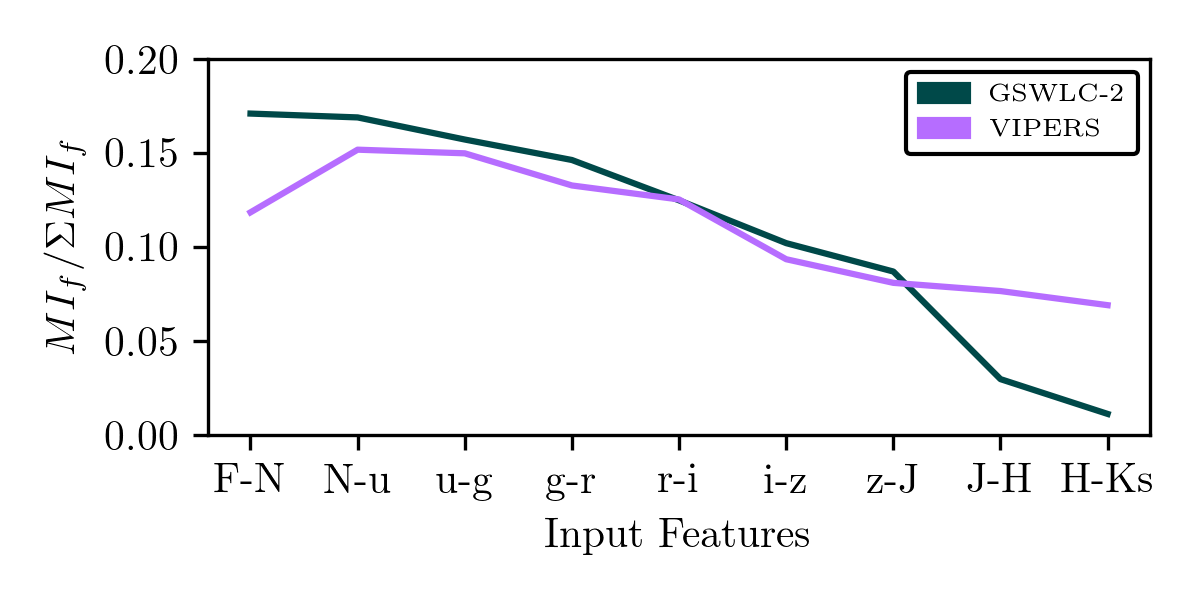}
\caption{The relative importance of each of the input features to the clustering. `F' stands for FUV, and `N' for NUV. The mutual information (see Section \ref{subsec:fi} and Equation \ref{eq:mi}) of each of the input features with respect to the cluster labels has been normalised by the sum across all of the input features for each sample.}
\label{fig:imp}
\end{figure}

In Fig. \ref{fig:imp}, we show the relative importance of each input feature to the clustering. Specifically, we calculate the mutual information ($MI$) between each input feature and the output cluster labels:

\begin{equation}
MI(f,l) = D_{KL}(p_{f,l} || p_{f} p_{l}).
\label{eq:mi}
\end{equation}

Here, $D_{KL}$ is the Kullback-Leibler divergence (\citealt{KULLBACK+1951}; also known as the relative entropy) between the joint probability distribution of input feature $f$ and output label $l$, and their independent distributions. For Fig. \ref{fig:imp}, $MI_{f,l}$ is normalised by its sum across all input features to give a relative value.

The lines in Fig. \ref{fig:imp} are broadly similar, indicating that, on the whole, \texttt{FEM} uses the nine features in a similar way to determine its best partitions. This is further confirmed by noting that the subspaces within which \texttt{FEM} determined these best partitions have the same dimensionality ($6$) for both samples. The lines are especially consistent among the optical colours, which is expected given that optical photometry is ubiquitously available for galaxies in both samples. Altogether, the optical regime is the most important to the clustering. Individually, colours from the UV region of the SEDs of the galaxies in both samples are most strongly related to the output cluster labels. This highlights, as expected, the star formation activity and the dust content of galaxies as major influences on the shapes of their UV-through-NIR SEDs.

UV colours are slightly more important for the clustering in our GSWLC-2 sample, which reflects the increased UV coverage of its galaxies by GALEX ($80$ per cent, as opposed to $52$ per cent for our VIPERS sample). NIR colours are less important for distinguishing clusters within our GSWLC-2 sample than within our VIPERS sample, which is likely due to their having been inferred purely from UV and optical input photometry. This is in contrast with the galaxies in our VIPERS sample, whose NIR SEDs (more important to the clustering) were instead constrained by $K_{s}$-band photometry\footnote{While the Two Micron All-Sky Survey \citep{SKRUTSKIE+2006} has NIR photometry for $\sim 50$ per cent of GSWLC-2 galaxies, it is shallow and would not have provided strong constraints upon their NIR SEDs.}. For galaxies with incomplete photometry, the array of templates and synthetic spectra with which \texttt{CIGALE} may fit them is reduced, leading to reduced variation in the shapes of their SEDs. In addition, the rest-frame magnitudes (and hence, rest-frame colours) that CIGALE must infer from photometry at other wavelengths have larger uncertainties. Hence, availability of photometry with which to constrain the SEDs of galaxies is advantageous to the clustering. Nevertheless, Fig. \ref{fig:imp} shows that, for the most part, \texttt{FEM} uses the features similarly to model both samples despite slight differences in this availability, which is driven mostly by the ubiquitous availability of optical photometry for both samples.

\begin{table*}
\caption{Profiles, in terms of averages, of the clusters determined within each of our samples. See the main text for an explanation of the cluster naming scheme. We list cluster means in columns $NUV-r$ and $r-K_{s}$. For the remaining features, which are less directly linked to the clustering, we opt for medians to mitigate the potential influence of outliers on the cluster profiles. Column `\%' lists the percentage of galaxies contained within each cluster for each sample. The data in the next seven columns [$NUV-r$ to $\log_{10}(sSFR/$yr$^{-1})$ (SED)] originates from the same \texttt{CIGALE} SEDs as the rest-frame colours that were used as inputs to the clustering. Features listed in this table include colour excesses [$E(B-V)$], stellar masses ($M_{*}$), stellar metallicities ($Z$), mass-weighted stellar ages ($MWSA$), and specific star formation rates ($sSFR$). We list sSFRs both determined by \texttt{CIGALE} (SED; averaged over $100$ Myr timescales) and determined from galaxy spectra (and hence independent of \texttt{CIGALE}; ind.; see Sections \ref{subsubsec:finalg} and \ref{subsubsec:finalv}). Medians marked with asterisks have unexpected values given their corresponding $NUV-r$ colour and are discussed in Section \ref{subsubsec:pass}.}
\label{tab:c}
\centering
\begin{tabular}{l r r r r r r r r r r}
\hline
Cluster & \% & $NUV-r$ & $r-K_{s}$ & $E(B-V)$ & $\log_{10}(M_{*}/$M$_{\odot})$ & $\log_{10}(Z)$  & $\log_{10}(MWSA/$Myr$)$ & \multicolumn{2}{c}{$\log_{10}(sSFR/$yr$^{-1})$} \\
 & & & & & & & & (SED) & (ind.) \\
\hline
G$1$ & $24.0$ & $2.39$ & $0.42$ & $0.11$  &  $9.90$ & $-2.22$  & $3.80$ &  $-9.87$ &  $-9.87$  \\
G$2$ & $15.2$ & $3.29$ & $0.91$ & $0.20$  & $10.26$ & $-1.81$  & $3.85$ & $-10.02$ & $-10.19$  \\
G$3$ & $17.3$ & $3.51$ & $0.78$ & $0.14$  & $10.37$ & $-2.11$  & $3.89$ & $-10.38$ & $-10.47$  \\
G$4$ &  $8.5$ & $4.31$ & $1.16$ & $0.13$  & $10.70$ & $-1.75$  & $3.92$ & $-10.87$ & $-11.22$  \\
G$5$ &  $9.7$ & $5.07$ & $0.67$ & $0.22$  & $10.35$ & $-2.30$  & $3.90$ & $-10.78$ & $-11.97$  \\
G$6$ & $11.3$ & $5.24$ & $0.78$ & $0.08$  & $10.57$ & $-2.11$  & $3.93$ & $-11.92$ & $-11.93$  \\
G$7$ & $14.0$ & $5.27$ & $0.73$ & $0.11$  & $10.54$ & $-2.20$  & $3.93$ & $-11.85$ & $-12.02$  \\
\hline
V$1$ & $26.8$ & $1.86$ & $0.25$ &  $0.01$ &  $9.87$ &  $-2.12$ &  $3.52$ &  $-9.34$ &  $-9.25$  \\
V$2$ & $18.4$ & $2.17$ & $0.60$ &  $0.02$ & $10.14$ &  $-1.90$ &  $3.55$ &  $-9.22$ &  $-9.34$  \\
V$3$ &  $9.3$ & $2.62$ & $0.75$ &  $0.05$ & $10.10$ &  $-1.40$ &  $3.52$ &  $-8.99$ &  $-9.35$  \\
V$4$ & $18.5$ & $3.26$ & $1.05$ &  $0.12$ & $10.67$ &  $-1.80$ &  $3.58$ &  $-9.71$ &  $-9.92$  \\
V$5$ &  $5.2$ & $4.75$ & $0.91$ & *$0.15$ & $10.61$ & *$-1.51$ & *$3.52$ & *$-9.43$ & $-10.09$  \\
V$6$ & $10.3$ & $4.81$ & $0.90$ & *$0.15$ & $10.69$ & *$-1.86$ & *$3.61$ & *$-9.90$ & $-10.29$  \\
V$7$ & $11.5$ & $4.86$ & $0.96$ &  $0.02$ & $10.91$ &  $-2.05$ &  $3.74$ & $-11.27$ & $-10.42$  \\
\hline
\end{tabular}
\end{table*}

\subsection{Clustering structures}
\label{subsec:struc}

Table \ref{tab:c} profiles the clusters determined within both samples. Features are derived from the same SEDs as the colours used for the clustering (see Sections \ref{subsubsec:gseds} and \ref{subsubsec:vseds}) as well as from ancillary sources (see Sections \ref{subsubsec:finalg} and \ref{subsubsec:finalv}). Clusters are named using two-part notation that will be used throughout the remainder of this paper; prefixes `G' or `V' denote clusters determined within the GSWLC-2 and VIPERS samples respectively. Clusters names have been ordered by their mean $NUV-r$ colours for ease of reference.

\begin{figure*}
\centering
\includegraphics[width=0.9\textwidth]{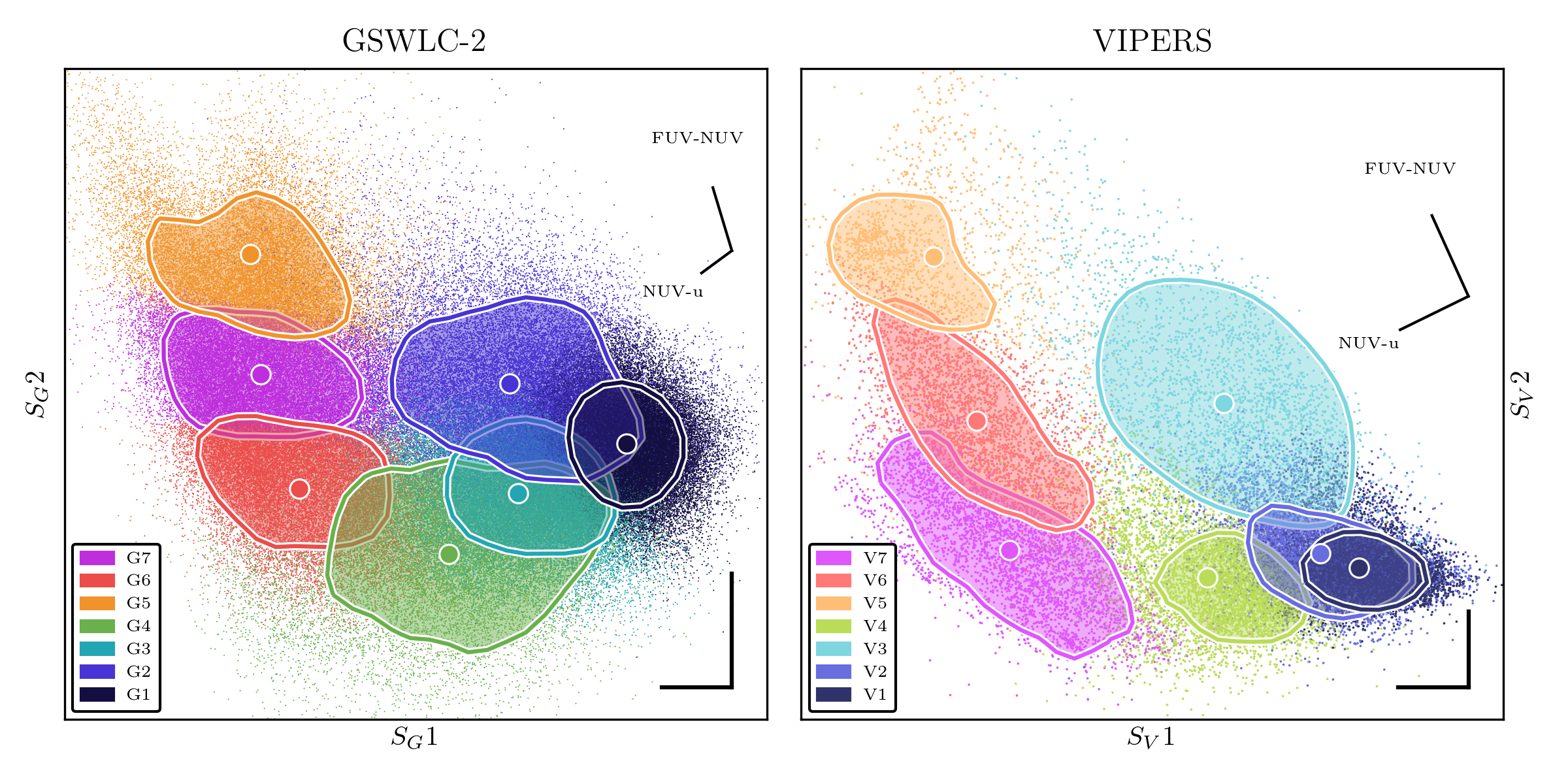}
\caption{Projections of both samples onto the two dimensions that best separate their clusters. The axes of each plot are determined by \texttt{FEM} and are unique to each sample (as indicated by their labels; e.g. $S_{G}1$ represents the first axis of the subspace of our GSWLC-2 sample). but the resultant projections are mostly similar nonetheless. The distributions of clusters within this plane are shown using coloured, filled contours (drawn at a relative density of $0.4$), and the coloured, circular markers show their means. The perpendicular black lines at the lower right of each plot show the extent to which the y-axis has been stretched relative to the x-axis to yield the projections as shown. The vectors at the upper right of each plot show the projections of the two input features that correlate most strongly with the axes of these projections.}
\label{fig:sub}
\end{figure*}

Fig. \ref{fig:sub} shows projections of our samples onto the two principal dimensions of their respective six-dimensional discriminative subspaces. These projections, which offer direct views of the structures of the clustering outcomes, are determined uniquely for each sample by \texttt{FEM}: \textit{the axes of the two plots do not correspond exactly to one another}. Nevertheless, these projections are broadly similar in terms of the shapes of the overall samples within them. Both samples exhibit continua in these projections, running from the lower right to the upper left of each plot, which have been segmented by \texttt{FEM}. That this segmentation is robustly reproducible over successive runs of \texttt{FEM} (Table \ref{tab:icl}) indicates that \texttt{FEM} has captured astrophysically meaningful structures in the samples. In addition, both samples exhibit a cluster which extends into the sparser region to the upper right of each plot. This overall similarity suggests that the evolution of galaxies at the epochs of the two samples is mostly similar. It also gives us confidence in the success of the measures taken to ensure a fair comparison between samples at different redshifts and from different surveys (see Sections \ref{subsubsec:finalg} and \ref{subsubsec:finalv}), and reinforces our conclusion that \texttt{FEM} has overall used the input features similarly for both samples in spite of slight differences in the availability of photometry between them (see Section \ref{subsec:fi}). The subtler differences between clusters in these projections are subject to the distributions of galaxies \textit{within} the shapes of their respective samples. We comment on these differences where relevant in Section \ref{subsec:ids}. Cluster colours in the plots in this paper, like their names, are assigned based on their mean $NUV-r$ colours.

We break down the analysis of our clusters by using the two-dimensional colour bimodality of galaxies as a simple framing device. The colour bimodality is a steady property of the galaxy population throughout cosmic time, having been observed among galaxies with redshifts as high as $4$ \citep{WUYTS+2007,WILLIAMS+2009,ILBERT+2010,ILBERT+2013}. Hence, we may use it to separate clusters that are more strongly associated with the blue peak (containing mostly star-forming galaxies) from clusters that are more strongly associated with the red peak (containing mostly passive galaxies) in a way that is independent of redshift.

\begin{figure*}
\centering
\includegraphics[width=0.76\textwidth]{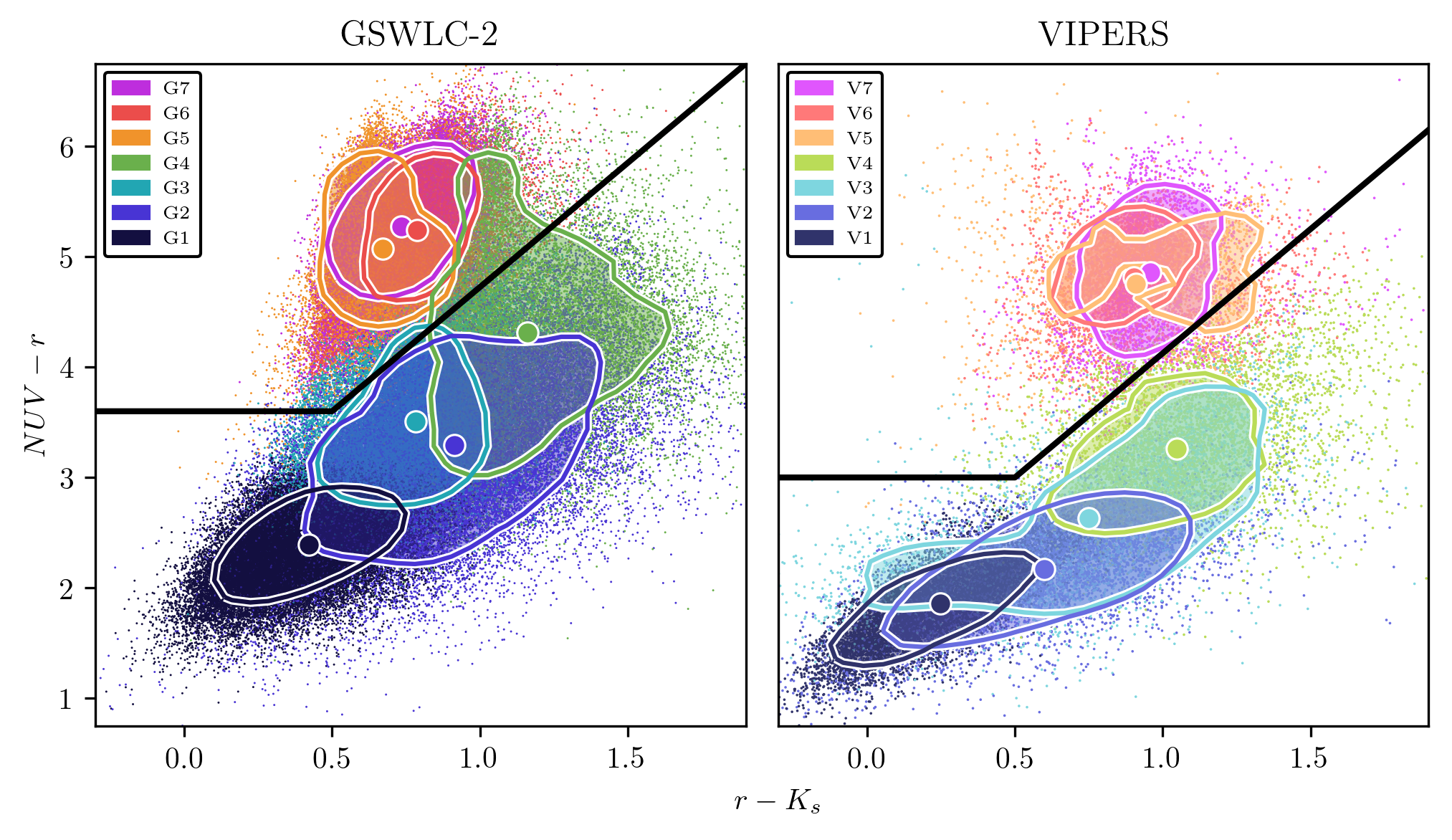}
\caption{Colour-colour plots of our samples. Colours are derived from \texttt{CIGALE} SED estimation. The distributions of clusters are shown using coloured, filled contours (drawn at a relative density of $0.4$), and the coloured, circular markers show their means. The black line in each plot (inspired by \citealt{MOUTARD+2016b}; see main text) marks the boundary between star-forming galaxies (below the line) and passive galaxies (above the line).}
\label{fig:nuvrks}
\end{figure*}

This two-dimensional separation is marked by the black lines in Fig. \ref{fig:nuvrks}. The $NUV-r-K_{s}$ colour-colour plane \citep{ARNOUTS+2013,MOUTARD+2016b} is a useful tool with which to probe galaxy subpopulations due to its ability to separate star-forming (low $NUV-r$), passive (high $NUV-r$), and also dusty (high $r-K_{s}$) galaxies. It has been applied in several studies of galaxy evolution using data from VIPERS (e.g. \citealt{FRITZ+2014,DAVIDZON+2016,MOUTARD+2016b,SIUDEK+2017,SIUDEK+2018a,VERGANI+2018}). The form of the black lines is inspired by \cite{FRITZ+2014} and \cite{MOUTARD+2016b}; they are placed independently in each panel, without reference to the positions of the clusters, to simply demarcate the star-forming and passive regions of the $NUV-r-K_{s}$ plane. Clusters whose means then lie below the black line in each plot are selected as `blue', `star-forming' clusters, and clusters whose means then lie above the black lines are selected as `red', `passive' clusters. As a result, both samples break down into four blue clusters and three red clusters. Deviations of the structures of the clusters from this simple blue/red (star-forming/passive) division that we enforce (e.g. clusters that overlap or span this division) will highlight limitations of a purely two-dimensional view of the galaxy population and its bimodality. The separation between these two main families of clusters suggests differences in the evolution processes influencing the galaxies that they contain.

The blue peak of the bimodality corresponds closely with the star-forming main sequence (SFMS; \citealt{NOESKE+2007,SALIM+2007}), which is the tight correlation between the SFRs and the stellar masses of actively star-forming galaxies. The SFMS, like the bimodality, is ubiquitous throughout cosmic time \citep{SPEAGLE+2014}. It has a lower normalisation with decreasing redshift; this cosmological decline of star formation \citep{MADAU+1996,MADAU+2014,DRIVER+2018} is visible as a vertical offset between the samples in Fig. \ref{fig:nuvrks}. In this paper, the terms `blue peak' and `SFMS' are synonymous, and we use them interchangeably.

The stronger $NUV-r$ split between star-forming and passive VIPERS clusters in comparison with those of GSWLC-2 (Figs. \ref{fig:nuvrks} and \ref{fig:sub}) is likely to result from two factors. First is the difference in the rest-frame wavelength coverage of GALEX photometry for the two samples; some rest-frame UV emission is redshifted out of the bandwidths of GALEX's filters at $z \sim 0.65$. Second is the difference in the completeness of UV photometry for each sample. GALEX observations exist for $\sim 80$ per cent of galaxies in clusters G1-4. This proportion falls to $\sim 55$ per cent in clusters G5-7, but this is expected given that these galaxies would be fainter in the UV regime. Meanwhile, $\sim 65$ per cent of V1, V2, and V4 galaxies were observed by GALEX. Interestingly, only $\sim 20$ per cent of galaxies in V3 have observed UV photometry, which may explain its separation from the other star-forming VIPERS clusters. Passive VIPERS clusters are $\sim 25$ per cent complete in observed UV photometry. Together, these factors mean we are likely to miss low levels of UV emission from more evolved VIPERS galaxies with more intermediate colours. On the other hand, Fig. \ref{fig:imp} shows that rest-frame $NUV-u$ colours are similarly important to the clustering structures of \textit{both} samples, with $NUV$ emission expected to be a particularly accurate tracer of star formation \citep{SALIM+2014}.

\subsection{Cluster identities}
\label{subsec:ids}

\begin{table}
\caption{Profiles, in terms of averages of ancillary features, of the clusters determined within each of our samples. See the main text for an explanation of the cluster naming scheme. We list the median values of the galaxies that the clusters contain for each of the features. Column `\%' lists the percentage of galaxies contained within each cluster for each sample. Features listed in this table include S\'ersic indices ($n_{g}$), half-light radii ($R_{1/2}$), and environmental overdensities ($\delta$). The data is drawn from ancillary sources (see Sections \ref{subsubsec:finalg} and \ref{subsubsec:finalv}).}
\label{tab:c2}
\centering
\begin{tabular}{l r r r r r r r r r r}
\hline
Cluster	& \% & $n_{g}$ & $\log_{10}(R_{1/2}/$kpc$)$ & $\log_{10}(1 + \delta)$ \\
\hline
G$1$ & $24.0$ & $1.04$ & $0.57$ & $0.40$ \\
G$2$ & $15.2$ & $1.34$ & $0.50$ & $0.51$ \\
G$3$ & $17.3$ & $1.57$ & $0.55$ & $0.55$ \\
G$4$ &  $8.5$ & $2.38$ & $0.61$ & $0.59$ \\
G$5$ &  $9.7$ & $4.09$ & $0.40$ & $0.85$ \\
G$6$ & $11.3$ & $4.18$ & $0.45$ & $0.80$ \\
G$7$ & $14.0$ & $4.25$ & $0.44$ & $0.83$ \\
\hline
V$1$ & $26.8$ & $0.92$ & $0.49$ & $0.29$ \\
V$2$ & $18.4$ & $0.95$ & $0.48$ & $0.29$ \\
V$3$ &  $9.3$ & $1.11$ & $0.50$ & $0.36$ \\
V$4$ & $18.5$ & $1.53$ & $0.55$ & $0.35$ \\
V$5$ &  $5.2$ & $3.31$ & $0.42$ & $0.40$ \\
V$6$ & $10.3$ & $3.29$ & $0.40$ & $0.40$ \\
V$7$ & $11.5$ & $3.40$ & $0.43$ & $0.43$ \\
\hline
\end{tabular}
\end{table}

\subsubsection{Clusters of star-forming galaxies}
\label{subsubsec:sfms}

\begin{figure}
\centering
\includegraphics[width=0.45\textwidth]{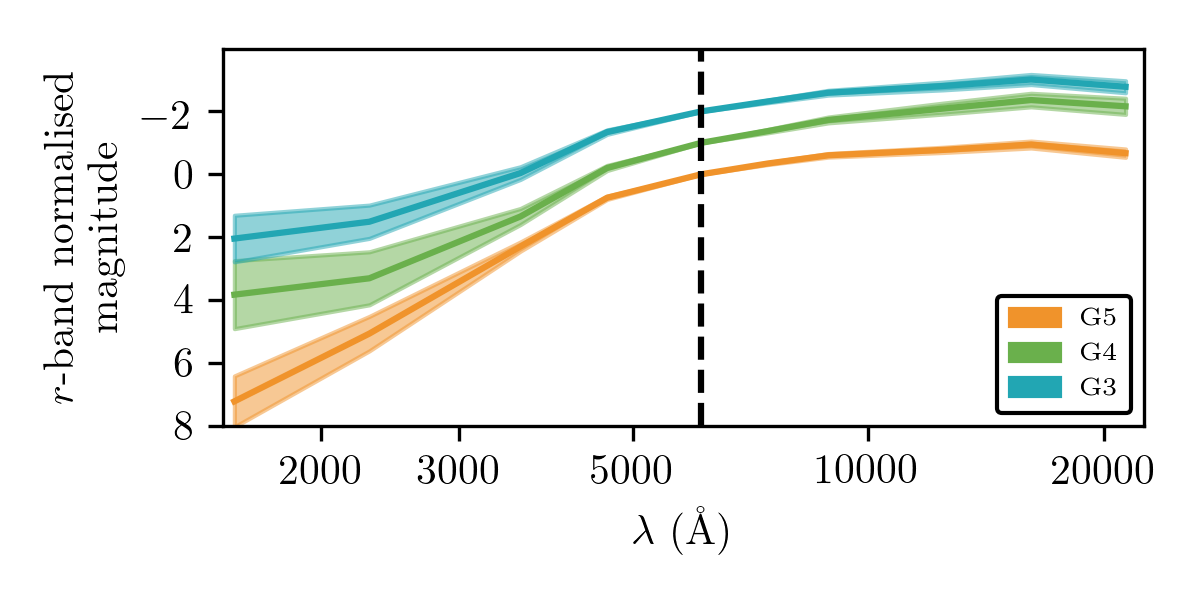}
\caption{A comparison of the shapes of the mean ($\pm$ standard deviation) estimated SEDs of galaxies in clusters G3, G4, and G5. Clusters G3 and G5 are chosen as they neighbour G4 in terms of their average $NUV-r$ colour. The estimated SEDs of individual galaxies are normalised by their $r$-band magnitudes (the effective wavelength of which is marked by a dashed black line) before the mean estimated SEDs are calculated. The y-axis applies to the mean SED of G5; those of G3 and G4 are vertically offset by $-1$ and $-2$ respectively to more clearly show the differences in their shapes.}
\label{fig:g4_seds}
\end{figure}

Our $NUV-r-K_{s}$ cut (Section \ref{subsec:struc}) yields the following blue clusters: G1, G2, G3, and G4 for the GSWLC-2 sample; and V1, V2, V3, and V4 for the VIPERS sample. While dominated by blue galaxies, clusters G4 and V4 also contain a significant number of galaxies with green or red $NUV-r$ colours (including the vast majority of green valley galaxies). Fig. \ref{fig:g4_seds} shows that the SEDs of G4 galaxies are generally more similar to those of actively star-forming galaxies, being flatter in the UV regime (e.g. G3 galaxies) than those of typically passive galaxies (e.g. G5 galaxies). Hence, in terms of the influence of their evolutionary histories on the shapes of their SEDs, G4 galaxies appear more closely related to G1-3 galaxies than to G5-7 galaxies, despite some G4 galaxies occupying the passive region of the $NUV-r-K_{s}$ plane in Fig. \ref{fig:nuvrks}. Similarly, the SEDs of V4 galaxies more closely resemble those of V1-3 galaxies rather than V5-7 galaxies (not shown).

Given that the SFMS is a smooth continuum, it is important where possible to establish why \texttt{FEM} has distinguished clusters within it, and to interpret the significance of these distinctions in terms of galaxy evolution. The position of a galaxy along the $NUV-r-K_{s}$ SFMS (Fig. \ref{fig:nuvrks}) is governed by a combination of its stellar mass and its dust content \citep{MOUTARD+2016a,MOUTARD+2016b}. The lobe at high $r-K_{s}$, which preferentially consists of edge on galaxies, is known to capture the excess reddening of high-mass star-forming galaxies \citep{ARNOUTS+2013}, but it is more difficult to disentangle this combination of stellar mass and dust elsewhere within the SFMS. Hence, we see an overlap of star-forming clusters in Fig. \ref{fig:nuvrks}. In Fig. \ref{fig:sub}, though, these clusters are more clearly separated.

G1 and V1 capture equivalent subpopulations of galaxies. Both clusters contain the galaxies with the bluest colours and the lowest masses (Fig. \ref{fig:nuvrks}, Table \ref{tab:c}) within their respective samples; star-forming galaxies at relatively early stages of their evolution. The remaining star-forming clusters have higher masses and lie further along the SFMSs of each sample.

Clusters G2 and G3 overlap with one another in the left-panel of Fig. \ref{fig:nuvrks}, as do clusters V2 and V3 in the right-hand panel of the same figure. Fig. \ref{fig:sub} shows that G2 and V3 both extend away from the main continua within the subspace projections of their respective samples. The feature vector projections in Fig. \ref{fig:sub} show that the galaxies in these clusters have particularly red $FUV-NUV$ colours in comparison with other SFMS clusters. However, the astrophysical meaning behind this is unclear. \texttt{CIGALE} alternately attributes this reddening to high colour excesses for galaxies in G2 and to higher metallicities for galaxies in V3 (Table \ref{tab:c}), suggesting that it has not fully resolved the degeneracy between the influences of dust and metallicity upon the colours of these galaxies. However, \texttt{CIGALE} \textit{is} consistent in assigning G2 and V3 galaxies similar stellar masses and mass-weighted stellar ages to G3 and V2 galaxies (Table \ref{tab:c}), which occupy similar regions of the $NUV-r-K_{s}$ plane. Stellar mass estimates are not strongly affected by an inability to resolve this degeneracy between the influences of dust and metallicity (e.g. \citealt{BELL+2001}). Clusters G3 and V2, lying on the main continua in Fig. \ref{fig:sub}, seem to be intermediate between clusters G1 and G4, and V1 and V4 respectively.

\begin{figure}
\centering
\includegraphics[width=0.39\textwidth]{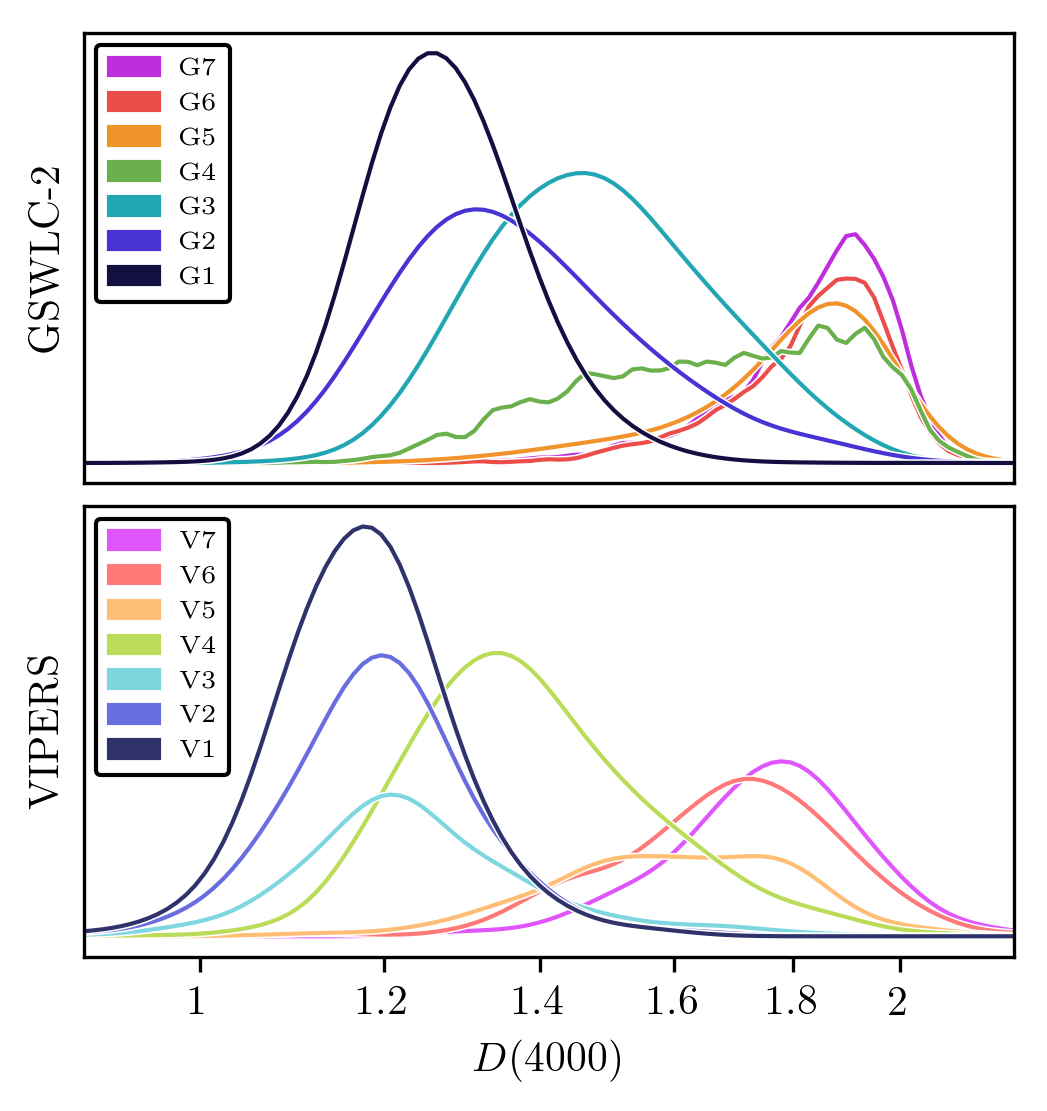}
\caption{Smoothed kernel density estimates in $D(4000)$ (logarithmically distributed) for each of the clusters from both outcomes. Here, $D(4000)$ was measured from the spectra of galaxies \protect\citep{BRINCHMANN+2004,GARILLI+2014} using a method introduced by \protect\cite{BALOGH+1999}, and is hence independent of \texttt{CIGALE}'s estimated SEDs.}
\label{fig:d4000}
\end{figure}

The star-forming clusters along the SFMS of our GSWLC-2 sample exhibit a gradient in their star formation activity. Taking their increasing average stellar masses as a point of reference, clusters G1-4 exhibit a corresponding increase in their average $NUV-r$ colours (Table \ref{tab:c}, Fig. \ref{fig:nuvrks}). decrease in their average sSFRs (both SED and ind.; Table \ref{tab:c}), and increase in their average $D(4000)$ (Fig. \ref{fig:d4000}). High-mass galaxies in our GSWLC-2 sample do not form stars as readily as low-mass galaxies. This gradient is weaker for clusters V1-3 (particularly with regard to their median sSFRs; Table \ref{tab:c}), though we note that clusters V2 and V3 have lower average stellar masses than G2 and G3. It is only in V4 that we see a rise in average stellar mass accompanied by a decrease in average $sSFR$, and an increase in $D(4000)$.

\begin{figure*}
\centering
\includegraphics[width=0.77\textwidth]{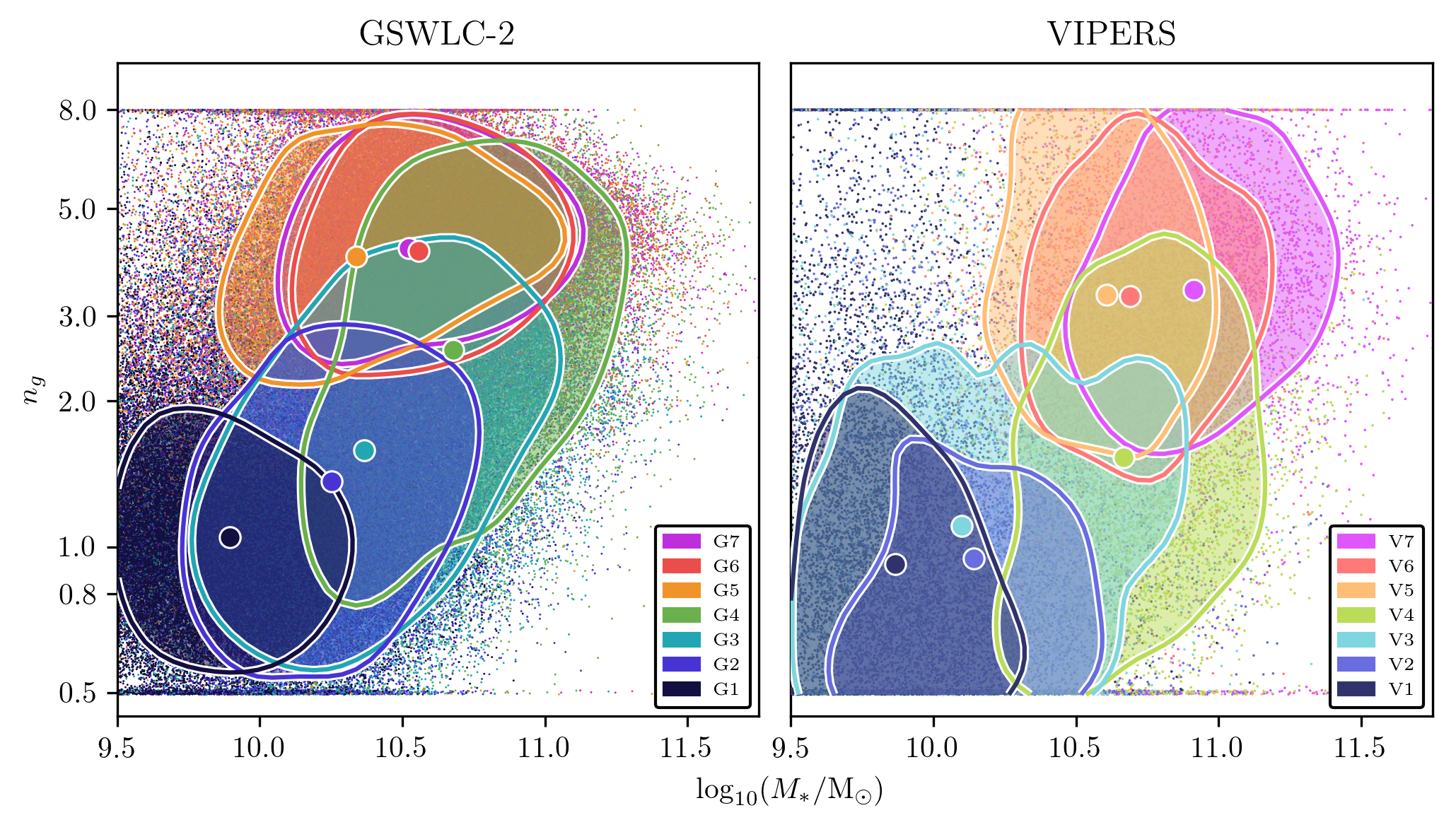}
\caption{S\'ersic index versus stellar mass for the galaxies in our samples. S\'ersic indices were determined by \protect\cite{SIMARD+2011} for our GSWLC-2 sample, and \protect\cite{KRYWULT+2017} for our VIPERS sample. The distributions of clusters are shown using coloured, filled contours (drawn at a relative density of $0.4$), and the coloured, circular markers show their medians. We have winsorised the S\'ersic indices of the galaxies in our VIPERS sample to values of $0.5$ and $8$ in order to match the limits of our GSWLC-2 sample.}
\label{fig:massng}
\end{figure*}

The large median sizes and low-to-intermediate median S\'ersic indices of star-forming clusters from both samples indicate that they are dominated by disc galaxies (Table \ref{tab:c2}). Clusters G1-4 exhibit a rise in their median $n_{g}$ to intermediate values along their SFMSs, indicating increasingly concentrated morphologies among their galaxies. In Fig. \ref{fig:massng}, these clusters form morphological sequences that are separate from the distributions of passive clusters in the same plane. The sequence of V1-4 is not as strong as that of G1-4; again, it is only in V4 that we see a significant change, with the higher stellar masses of its galaxies met with intermediate S\'ersic indices.

\begin{figure}
\centering
\includegraphics[width=0.39\textwidth]{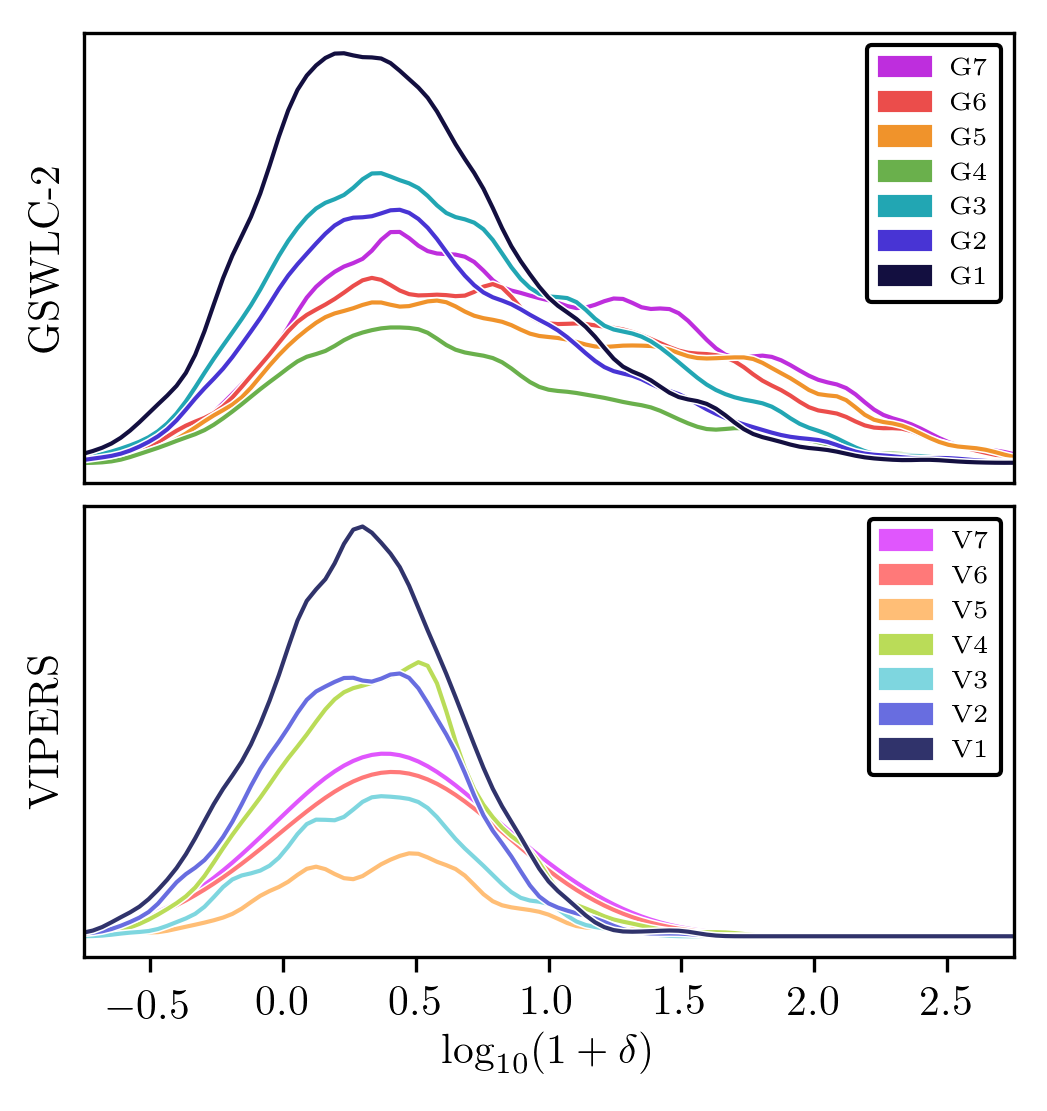}
\caption{Smoothed kernel density estimates in local environmental overdensity ($\delta$) for each of the clusters from both outcomes. For both samples, these overdensities are based on fifth-nearest neighbour surface densities \protect\citep{BALDRY+2006,CUCCIATI+2017}.}
\label{fig:env}
\end{figure}

While there are slight trends in the median local environmental overdensities of the star-forming clusters in both samples (Table \ref{tab:c2}), Fig. \ref{fig:env} shows that their distributions thereof have very large spreads and exhibit a great deal of overlap with the distributions of other SFMS clusters from the same sample. Therefore, we cannot attribute the reduction in the star formation activity of SFMS galaxies at higher masses to mainly environmental causes for either sample.

\subsubsection{Clusters of passive galaxies}
\label{subsubsec:pass}

Our red clusters, selected in two dimensions using the $NUV-r-K_{s}$ plots in Fig. \ref{fig:nuvrks}, are: G5, G6, and G7 for our GSWLC-2 sample, and V5, V6, and V7 for our VIPERS sample. The colour that best separates the passive clusters in both samples is $FUV-NUV$. For G5-7, this separation corresponds with the higher sSFRs and lower masses of G5 galaxies, and differences in the metallicities of G6 and G7 galaxies (Table \ref{tab:c}). V7 has been distinguished due to the high masses and low sSFRs of its galaxies. However, \texttt{CIGALE}'s estimation of the astrophysical properties of V5 and V6 galaxies is less reliable (see below). In general, galaxies in the passive clusters are offset to redder $NUV-u$ colours than those in the SFMS clusters (Section \ref{subsubsec:sfms}).

Galaxies in clusters G6, G7, and V7 are alike with respect to most features. They share high stellar masses, low sSFRs, large $D(4000)$ (Fig. \ref{fig:d4000}), and early-type morphologies (Table \ref{tab:c}), all of which are typical of canonically passive galaxies. \texttt{CIGALE} attributes the difference in the $FUV-NUV$ colours of G6 and G7 galaxies (i.e. the feature that best separates these clusters) to their metallicity distributions. While G6 peaks strongly at $Z \sim -2.1$, G7 is split evenly between peaks at $Z \sim 2.1$ and $Z \sim -2.4$. The metallicities of passive GSWLC-2 galaxies are discretised by the input \cite{BRUZUAL+2003} grid, and due to a lack of any input NIR photometry during their SED estimation (see Appendix \ref{app:sm}); with more precise metallicities, their distributions might overlap more. V7 also has low metallicities in comparison with other clusters determined in its sample. We note that these sub-solar metallicities are unexpected for high-mass passive galaxies (e.g \citealt{GALLAZZI+2006}), indicating difficulties of breaking the age-dust-metallicity degeneracy with photometry alone, and suggesting that these metallicities are not entirely reliable. Altogether though, these clusters contain the oldest, most evolved galaxies among their respective samples: a subpopulation that is in place at the epoch of our VIPERS sample.

Galaxies in cluster G5, while also passive and early-type, have lower stellar masses than those in clusters G6 and G7. We also note a difference in the G5 median sSFRs as reported by \texttt{CIGALE} (SED) and by the \cite{BRINCHMANN+2004} calibration (ind.; Table \ref{tab:c}). G5 may contain post-starburst galaxies (PSBs; \citealt{WILD+2009}), with this difference in sSFRs possibly arising due to the different timescales probed by these two measures (see Section 7 of \citealt{SALIM+2016}). While the fibre component of $sSFR$ (ind.) is a more instantaneous measure of star formation activity ($\sim 10$ Myr, based on H$\alpha$ emission), \texttt{CIGALE} averages star formation over a longer period of time ($100$ Myr, matching the timescale traced by UV emission). Hence, even if the tail of a declining central burst of star formation activity is not captured by $sSFR$ (ind.), it may still be captured by $sSFR$ (SED). The spheroidal morphologies (Fig. \ref{fig:massng}, Table \ref{tab:c2}) and enhanced local environmental densities of G5 galaxies suggest an external influence upon their evolution (see Section \ref{subsec:env}), which is consistent with previous studies which link PSBs with mergers \citep{ZABLUDOFF+1996,YANG+2008,ALMAINI+2017}.

\begin{figure}
\centering
\includegraphics[width=0.45\textwidth]{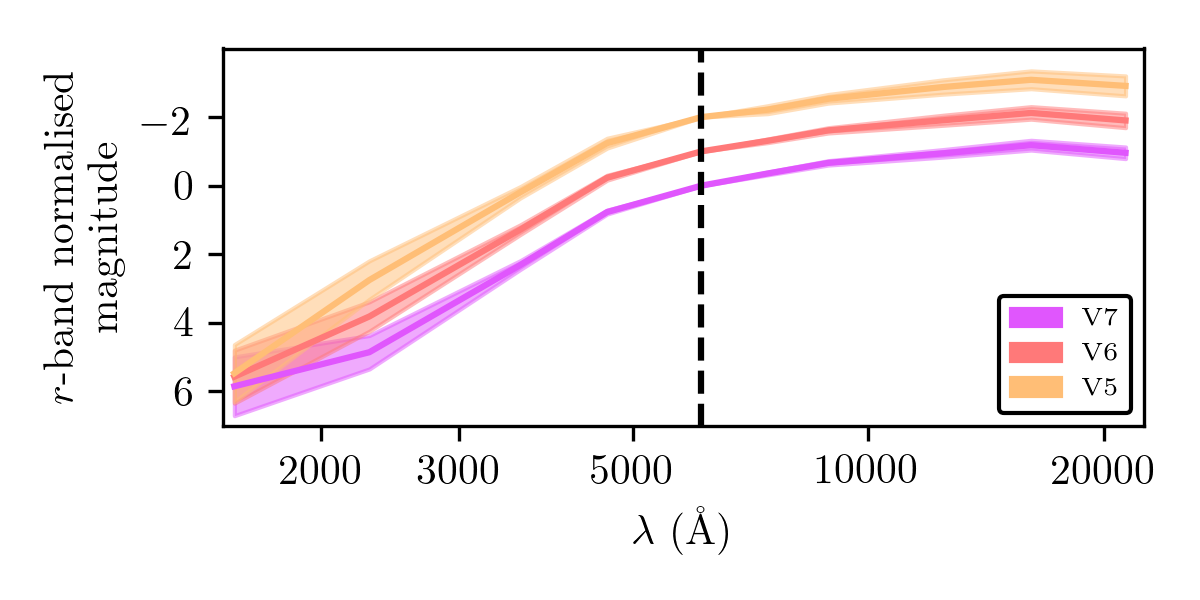}
\caption{A comparison of the shapes of the mean ($\pm$ standard deviation) estimated SEDs of galaxies in clusters V5, V6, and V6. The estimated SEDs of individual galaxies are normalised by their $r$-band magnitudes (the effective wavelength of which is marked by a dashed black line) before the mean estimated SEDs are calculated. The y-axis applies to the mean SED of V7; those of V5 and V6 are vertically offset by $-1$ and $-2$ respectively to more clearly show the differences in their shapes.}
\label{fig:red_seds}
\end{figure}

Clusters V5 and V6 present conflicting identities in terms of features estimated by \texttt{CIGALE} (Table \ref{tab:c}). While their galaxies have very similar stellar masses and morphologies to those in V7 (Table \ref{tab:c2}), they have unusually high colour excesses, metallicities, and $sSFR$ (SED). This is in contrast with the $sSFR$ (ind.) and observed $D(4000)$ values of these galaxies (Table \ref{tab:c}, Fig. \ref{fig:d4000}), which show that they are indeed passive. The large spread in $D(4000)$ of V5 may be due to some minor contamination of the cluster by star forming galaxies; its $NUV-r-K_{s}$ contour extends below the black line in Fig. \ref{fig:nuvrks}, into the region containing dusty star-forming galaxies. This may also drive its median $E(B-V)$ to a higher value.

The inability of \texttt{CIGALE} to properly resolve the age-dust-metallicity degeneracy for V5 and V6 galaxies is due to the UV regions of their SEDs. Fig. \ref{fig:red_seds} shows that V5 and V6 have steeper average UV SEDs than V7. To explain the red UV colours (especially $FUV-NUV$) of their galaxies, \texttt{CIGALE} invokes high colour excesses and metallicities rather than low $sSFR$ (SED). This appears to be a consequence of \texttt{CIGALE}'s two-burst SFHs, which may not be a realistic description of the SFHs of most passive VIPERS galaxies. These SFHs were adjusted for the epoch of our VIPERS sample by setting the formation time of the old population to $6.5$ Gyr ago instead of $10$ Gyr, and including the possibility of a particularly recent burst of star formation ($<50$ Myr). However, a trial of the use of a gradual $1$ Gyr quenching episode instead led to improvements in the quality of fit of passive SEDs (with low $sSFR$) to the photometry of the majority of V5 and V6 galaxies. Hence, it seems that further adjustments to \texttt{CIGALE}'s SFH prescription are required when applying it at higher redshifts\footnote{\texttt{LePhare} \citep{ILBERT+2006} SED estimation for the same galaxies \citep{MOUTARD+2016b,SIUDEK+2018a} used single exponentials for its SFHs and reported lower colour excesses, metallicities, and $sSFR$.}.

Galaxies contained within the passive clusters of our VIPERS sample tend to have higher stellar masses than those contained within the passive clusters of our GSWLC-2 sample (Table \ref{tab:c}). This is likely to be driven by stellar mass incompleteness of our VIPERS sample. \cite{DAVIDZON+2013} show that, even at its lower redshift limit of $z = 0.5$, our VIPERS sample is incomplete\footnote{Our `secure' redshift criterion (Section \ref{subsec:vipers}) may contribute slightly to this incompleteness (i.e. by selecting against faint, passive VIPERS galaxies that lack emission lines or strong absorption lines). However, \cite{DAVIDZON+2013} used a more relaxed criterion, so we do not expect our use of this criterion to significantly influence stellar mass completeness.} in passive galaxies below $\sim 10^{10}$ M$_{\odot}$. Furthermore, their completeness threshold increases with redshift to $10^{10.75}$ M$_{\odot}$ at our upper limit of $z = 0.8$, and thus skews our clusters of passive VIPERS galaxies towards higher stellar masses\footnote{Star-forming galaxies and clusters are affected to a much lesser degree.}. Hence, where the GSWLC-2 sample has two lobes of passive galaxies in Fig. \ref{fig:sub} (see also Appendix \ref{app:sm}), which differ in average stellar mass by $\sim 0.5$ dex, the VIPERS sample has only one. Though our VIPERS sample does contain some passive galaxies with low stellar masses (e.g. Fig. \ref{fig:massng}), they are not substantial enough in number for \texttt{FEM} to model them with a dedicated cluster (i.e. like G5).

Passive clusters in both samples have high S\'ersic indices and compact sizes (Table \ref{tab:c2}), indicating spheroid-dominated morphologies. They occupy separate regions of the plots in Fig. \ref{fig:massng} to their respective SFMS clusters. Fig. \ref{fig:massng} also shows that the $n_{g}$ distributions for passive clusters are highly consistent with one another. While the passive clusters in our GSWLC-2 sample exhibit a slight offset to higher density environments in comparison with star-forming GSWLC-2 clusters, the environments of passive VIPERS clusters are consistent with those of star-forming VIPERS clusters. This difference between the two samples is, in part, expected, due to the emergence of environments of especially high densities over cosmic time (e.g. \citealt{MARINONI+2008,KOVAC+2010,FOSSATI+2017}). However, factors such as spectroscopic fibre collisions and the aforementioned incompleteness of passive VIPERS galaxies may also reduce the completeness of VIPERS at high densities. This incompleteness does not appear to have strongly affected clusters elsewhere in the feature space (Fig. \ref{fig:sub}).

\section{Discussion}
\label{sec:disc}

Our clusters have been determined on the basis of the rest-frame colours of galaxies alone. In this section, we aim to discern what the trends of these purely colour-based clusters with other, ancillary features (see Sections \ref{subsubsec:sfms} and \ref{subsubsec:pass}) tell us about how strongly the SEDs of their constituent galaxies encode their evolution.

\subsection{Internally driven evolution}
\label{subsec:bulge}

\begin{figure}
\centering
\includegraphics[width=0.43\textwidth]{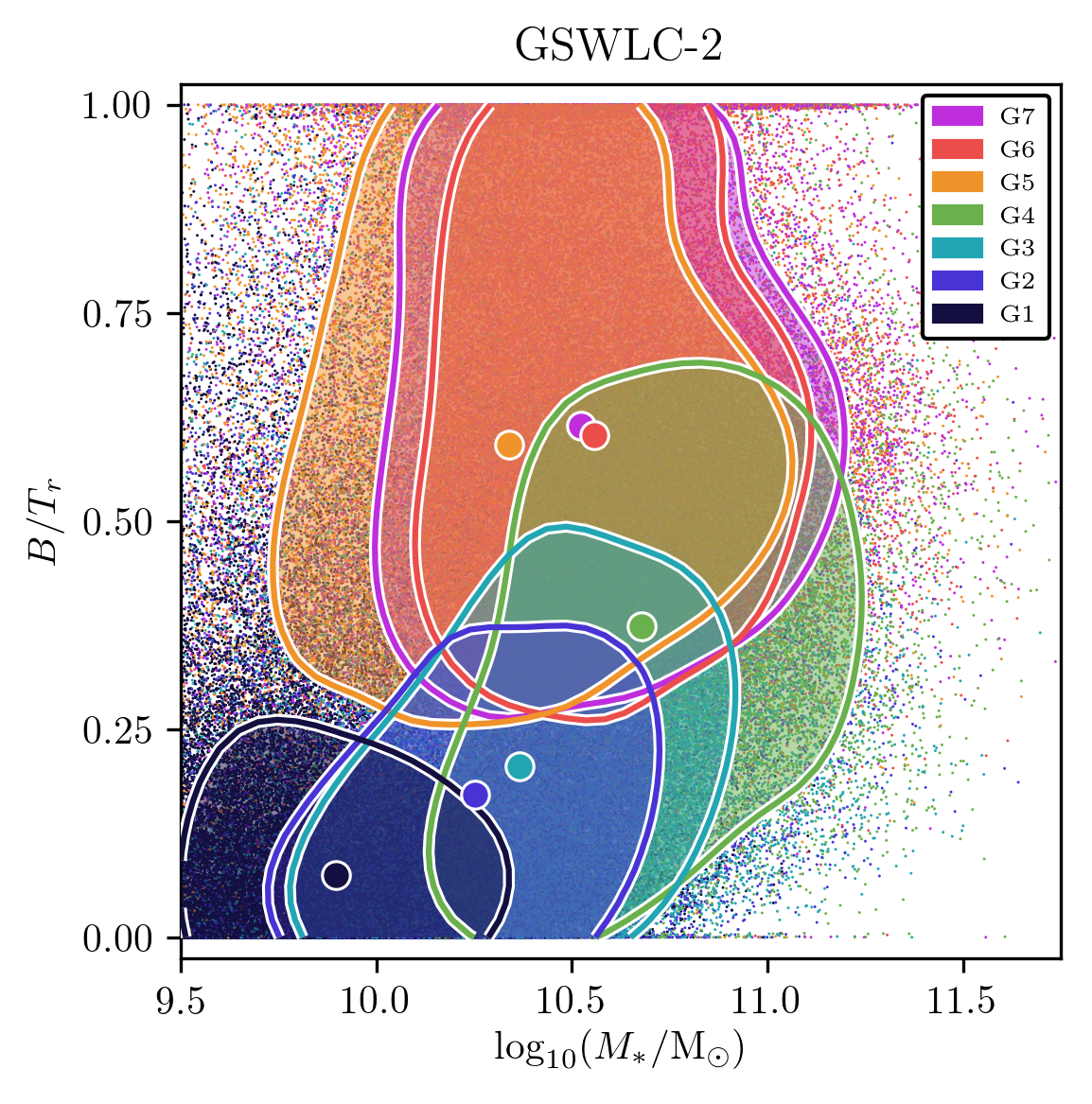}
\caption{Bulge-to-total ratio ($B/T_{r}$) versus stellar mass for the galaxies in our GSWLC-2 sample. Here, the subscript `$r$' denotes the $r$-band photometry from which the ratios were derived (\protect\citealt{SIMARD+2011}; based on two-component fits). The distributions of clusters are shown using coloured, filled contours (drawn at a relative density of $0.4$), and the coloured, circular markers show their means.}
\label{fig:bt}
\end{figure}

Alongside being closely related in terms of the shapes of their SEDs (see Section \ref{subsubsec:sfms}), both sets of star-forming clusters -- G1-4 and V1-4 -- form clear morphological sequences in Fig. \ref{fig:massng}. In Fig. \ref{fig:bt}, we examine the bulge-to-total ratios of GSWLC-2 galaxies using two-component \cite{SIMARD+2011} fits (no such data exists for VIPERS). The G1-4 sequence is apparent here as well, capturing the rising prominences of the bulges of their galaxies. It does \textit{not} extend to the highest $B/T_{r}$ values, despite G4 also containing some quenching and quenched galaxies. This indicates that G1-4 galaxies retain their discs as they evolve and that some G1-4 galaxies become passive without fully transforming their morphologies. The changing bulge-disc balance appears to be captured also in the large spread in $D(4000)$ of G4 galaxies in particular (Fig. \ref{fig:d4000}). The overlapping environmental distributions of star-forming clusters in both samples (Fig. \ref{fig:env}) suggest that these morphological sequences of gradual bulge growth are more likely to be due to internal processes (i.e. that act in all environments). We assume that our interpretation in this paragraph applies to galaxies in V1-4 as well.

Bar-driven inflows of star-forming gas \citep{SHETH+2005} -- an internal process that acts over long timescales -- constitute a likely candidate process. These inflows are commonly invoked to explain the formation of dynamically cold `pseudobulges' ($n_{cl} \lesssim 2$) rather than the dynamically hot `classical' ($n_{cl} \gtrsim 2$) bulges that the \cite{SIMARD+2011} two-component fits assume \citep{KORMENDY+2004,FISHER+2008,MISHRA+2017}. However, an increase in the prominence of pseudobulges would nonetheless be expected to be captured by the single-component fits which yield the S\'ersic indices in Table \ref{tab:c2} and Fig. \ref{fig:massng}. We do not rule out that SFMS galaxies may have undergone major and/or minor mergers or clump migration (a faster, more violent internal process; \citealt{ELMEGREEN+2008b,BOURNAUD+2011,TONINI+2016}) in their pasts; some have high total $n_{g}$ values, which may capture classical bulges formed as a result of these processes. Instead, we proffer that the processes do not contribute to the gradual of the bulges of these galaxies. It has been shown, for example, that the remnant of a gas-rich merger can reform a disc and continue to form stars, thus rejoining the SFMS \citep{HOPKINS+2009b,HOPKINS+2009a}.

The falling sSFRs of galaxies along the sequences G1-4 and V1-4 suggests that their morphologies are also linked with their quenching. This could be due to morphological quenching (i.e. the gravitational influence of the morphological components of galaxies upon star formation; \citealt{MARTIG+2009}). It is more likely, though, that the prominences of the bulges among these galaxies are a marker of nuclear activity. More massive bulges host more massive black holes at their centres \citep{HAERING+2004}, which supply more feedback energy to their surrounding galaxies. This feedback can inhibit further star formation by ejecting star-forming gas \citep{CROTON+2006,GABOR+2011,VERGANI+2018} or by preventing the cooling of newly accreted gas (above the `transition mass', $\sim 10^{10.5}$ M$_{\odot}$ at $z \sim 0$; \citealt{KAUFFMANN+2003b,DEKEL+2006,KERES+2009a,MOUTARD+2020}).

\begin{figure}
\centering
\includegraphics[width=0.45\textwidth]{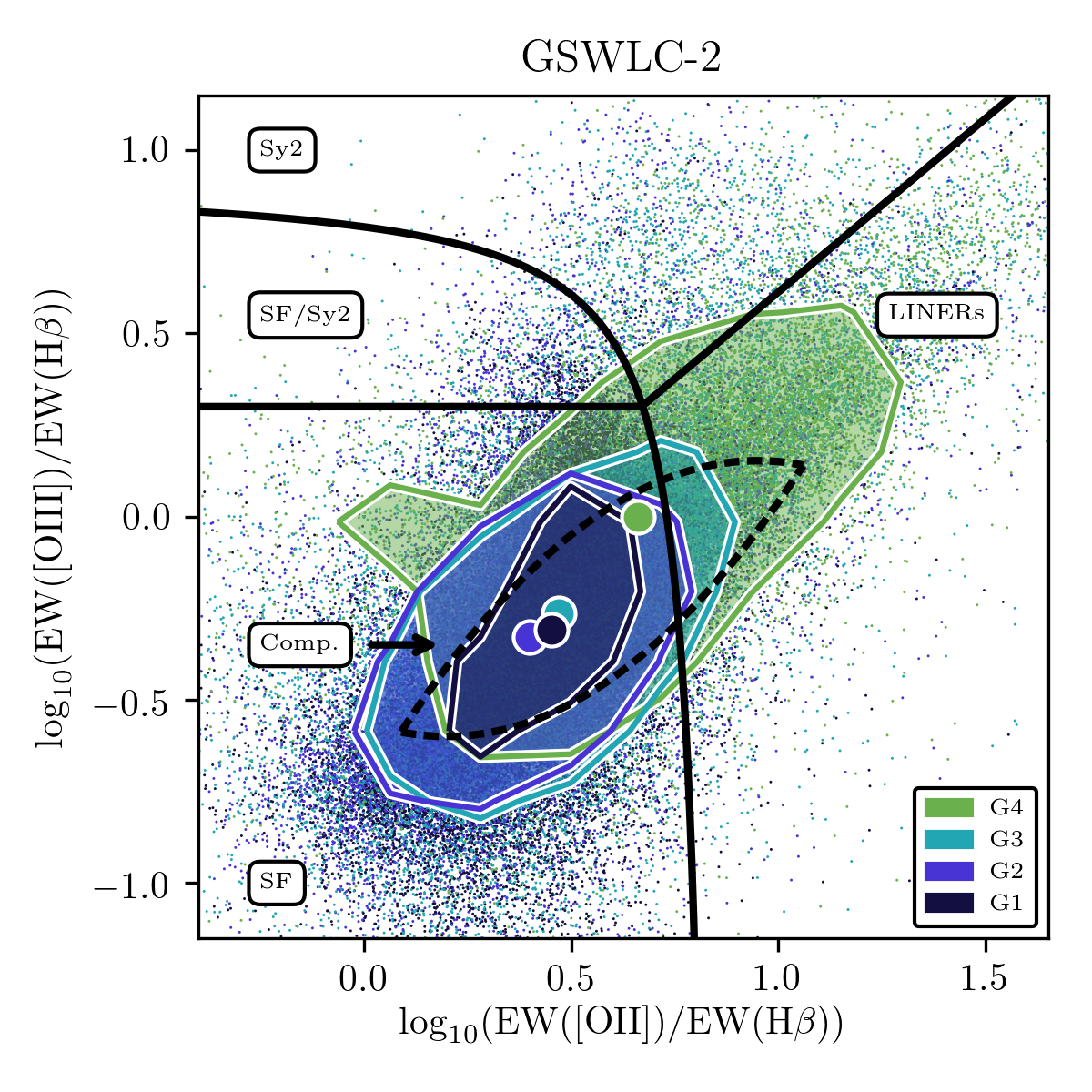}
\caption{A diagram for the classification of emission-line galaxies \citep{LAMAREILLE+2010} in our GSWLC-2 sample. Different regions, labelled and demarcated by black lines, correspond to different types of galaxy: `Sy2' to type II Seyfert galaxies, `SF' to purely star-forming galaxies, 'SF/Sy2' to a mixture of type II Seyfert and star-forming galaxies, `LINERs' to galaxies containing low-ionisation nuclear emission-line regions, and `Comp.' to a mixture of LINERs and star-forming galaxies. The distributions of clusters are shown using coloured, filled contours (drawn at a relative density of 0.4), and the coloured, circular markers show their medians.}
\label{fig:agn}
\end{figure}

Fig. \ref{fig:agn} shows the distributions of clusters G1-4 within the \cite{LAMAREILLE+2010} emission-line classification diagram. This diagram is chosen with a view to its applicability to galaxies at higher redshifts as well. The equivalent widths of the relevant emission lines were determined by \cite{BRINCHMANN+2004}, and are available for $94$ per cent of the galaxies in G1-4. Spectroscopy of these emission lines exists for some VIPERS galaxies as well \citep{GARILLI+2014}, but only for $34$ per cent of them, such that we would not be confident in the significance of any trend of our VIPERS clusters within the diagram. We note, however, that the few VIPERS galaxies for which this spectroscopy does exist tend to lie within the `SF' region of the plot, above the `Comp.' region (i.e. as in figure 10 of \citealt{SIUDEK+2018a}). Hence, we tentatively suggest a minimal influence of active galactic nuclei upon their current evolution, but reiterate that more data is needed to confirm this.

Clusters G1-4 are all centred in the `Comp.' region of Fig. \ref{fig:agn}, indicating that galactic nuclei are prevalent throughout them. G4 in particular extends well into the `LINERs' region of the diagram. Given the enhancement in the S\'ersic indices of G4 galaxies over G1-3 galaxies (Table \ref{tab:c2}, Fig. \ref{fig:massng}), this is consistent with previous studies which find that low-ionisation nuclear emission-line regions are more common in galaxies with earlier-type morphologies \citep{HECKMAN+1980}. In addition, this increase in nuclear activity for G4 galaxies coincides with their decrease in $sSFR$ in comparison with G1-3 galaxies (Table \ref{tab:c}), supporting the suggestion that supermassive black holes are involved in their quenching.

That the sSFRs of V1-3 galaxies do not decline as strongly as those of G1-3 galaxies may be tied to their morphologies; all three also have very low median $n_{g}$. This suggests that their bulges and/or supermassive black holes have not yet grown to the extent that they can effectively inhibit star formation. This would be consistent with \cite{FANG+2013} and \cite{BLUCK+2014}, who find that bulges must exceed a threshold in mass or central density before they become associated with quenching. For V4 galaxies, the reduction in sSFR is met with a rise to intermediate median $n_{g}$, suggesting that this threshold bulge mass has been achieved in some V4 galaxies.

Altogether, G1-4 and V1-4 galaxies (which include the vast majority of green valley galaxies) appear to evolve slowly and secularly \citep{SCHAWINSKI+2014,ILBERT+2015,MOUTARD+2016b,PACIFICI+2016a}. This is reflected in the similarity of their SEDs, which all feature relatively flat UV regions that suggest a gradual reduction in their star formation over time. It is also reflected in the morphological sequences that their clusters exhibit. The rising bulge prominences and declining star formation rates of these galaxies suggests that nuclear feedback, fuelled by bar-driven inflows, is the main mechanism driving their evolution \citep{GABOR+2011,MOUTARD+2020}. While this mechanism appears to act at the epochs of both samples, the connection between morphologies and star formation is stronger at lower redshifts. This may be linked with the long timescales over which these internal processes act, such that the gradual evolution of V1-4 galaxies may eventually lead to the more evolved distribution of galaxies given by clusters G1-4, which we assume to be their descendants. Hence, the rising prevalence of bulges grown by internal processes over cosmic time (e.g. \citealt{BRUCE+2012,GU+2019}) would appear to be linked to the cosmic decline of cosmic star formation activity. This connection between the bulges and the star formation of SFMS galaxies has previously been established \citep{CHEUNG+2012,FANG+2013,BLUCK+2014,CANODIAZ+2019,MCPARTLAND+2019}, but in our case it emerges purely from our clustering of galaxy colours, with morphologies invoked post-clustering for interpretation. Our clustering also appears to demonstrate that the SFMS is a two-dimensional projection of this pathway which, in the full nine-dimensional colour space, extends continuously to also include high-mass passive galaxies (as revealed by G4 in particular) that retain their discs, but are degenerate with other passive galaxies in two dimensions.

\subsection{Satellite quenching at low redshifts}
\label{subsec:env}

The uniformly red $NUV-u$ colours and the uniformly high S\'ersic indices of galaxies in clusters G5-7 and V5-7 imply a strong link between their passiveness and their concentrated morphologies. At high masses, this link may be obfuscated by a contribution from the internally-driven evolutionary pathway that we propose in Section \ref{subsec:bulge}. We note that cluster V7 in particular, containing VIPERS galaxies with the highest masses, seems to align well with the sequence of clusters V1-4 in Fig. \ref{fig:massng}, such that it could partially be an extension of this evolutionary pathway consisting of the oldest galaxies with the most prominent bulges. This is in agreement with previous studies which find that the inner stellar density of galaxies is a successful predictor of its having been quenched \citep{CHEUNG+2012,FANG+2013,BLUCK+2014}.

However, other passive clusters are separated from their respective sequences of star-forming clusters in Fig. \ref{fig:massng}. Clusters G7, G6, and especially G5 (the latter containing the lowest-mass passive galaxies in our GSWLC-2 sample) have high median $n_{g}$ in comparison with other clusters centred at similar stellar masses (G2, G3). This separation invites the interpretation that their galaxies are subject to alternative or additional evolutionary processes. That these clusters contain those GSWLC-2 galaxies that occupy the highest-density environments (Fig. \ref{fig:env}) suggests an additional influence of external processes. Hence, we suspect that a significant proportion of galaxies among G5-7 are satellite galaxies (occupying the halos of more massive central galaxies; \citealt{ILBERT+2010,MUZZIN+2013,MOUTARD+2018}). There is a weaker morphological separation for V5-7, and no environmental offset, which we attribute mostly to the incompleteness of low-mass passive galaxies in our VIPERS sample; these would also be expected to trace high-density environments. Hence, our following discussion on the influence of external processes upon satellite galaxies is conducted with reference to G5-7 only. Fully establishing whether external processes influence the evolution of low-mass passive galaxies at $z \sim 0.65$ in the same way requires a more complete sample.

Major and minor mergers \citep{TOOMRE+1977,BARNES-1988,BARNES-1992,WALKER+1996} and harassment \citep{MOORE+1996,SMITH+2015}, more common in environments of higher densities \citep{RENZINI-1999,TONINI+2016}, are external processes which can increase the S\'ersic indices of galaxies by transforming their morphologies from disc- to spheroid-dominated \citep{NAAB+2006,ACEVES+2006,FISHER+2008}. Fig. \ref{fig:bt} shows a range of bulge-to-total ratios among galaxies in G5-7, which may be capturing the varying degrees to which these processes disrupt their morphologies. While most G5-7 galaxies are strongly spheroid-dominated, others (while still having high S\'ersic indices) retain a disc component (with $B/T_{r}$ values as low as $\sim 0.3$). Whether these processes are also responsible for the quenching of G5-7 galaxies is unclear. Gravitational interactions between merging galaxies can induce central starbursts which rapidly exhaust their supplies of star-forming gas (e.g. PSBs, which we suggest comprise G5), and/or can catalyse nuclear activity which inhibits further star formation \citep{MIHOS+1994a,MIHOS+1994b,MIHOS+1996,DIMATTEO+2005,SPRINGEL+2005b,SPRINGEL+2005a}. However, a sufficiently gas-rich major merger may instead lead its remnant to form with a disc and continue forming stars \citep{BARNES-2002,HOPKINS+2009b,HOPKINS+2009a,HOPKINS+2010}. In addition, a merger remnant may accrete new gas such that it can form a new disc and renew star formation \citep{SALIM+2010,GABOR+2011}. Generally, mergers cannot be unequivocally linked with the quenching of galaxies (see also \citealt{WEIGEL+2017}), and so it is more likely that galaxies are quenched mainly by other processes.

Several external processes have been proposed to explain the quenching of star-forming galaxies as they become satellites. Examples include ram-pressure stripping \citep{GUNN+1972,MCCARTHY+2008}, thermal evaporation \citep{COWIE+1977,NIPOTI+2007}, and viscous stripping \citep{NULSEN-1982,KRAFT+2017}, all of which invoke the removal of the cold interstellar medium of a galaxy via its hydrodynamical interaction with the hot intergalactic medium of high-density environments as the reason for quenching. These processes are correlated with the velocity of a galaxy as it travels through its environment, and generally quench galaxies quickly. Gas may also be removed from the extended halo of a galaxy at the outskirts of a dense environment, \textit{by} the gravitational influence of that environment as a whole (`strangulation' or `starvation'; \citealt{LARSON+1980,PENG+2015}). The galaxy then quenches slowly by exhausting any remaining gas in its disc. The balance of these processes is not yet known \citep{BAHE+2015,PENG+2015,SMETHURST+2017}, but recent studies advocate for a general `delayed-then-rapid' quenching pathway \citep{WETZEL+2012,WETZEL+2013,MUZZIN+2014,MOUTARD+2018}. Galaxies initially quench slowly at the outskirts of the environment, then quickly as they approach its core, where the conditions for the aforementioned hydrodynamical interactions are expected. This delay could also explain the large spreads in the environmental distributions among all of our clusters in Fig. \ref{fig:env}. These quenching processes are, in turn, unlikely to transform the morphologies of low-mass passive galaxies \citep{BEKKI+2002,BOSELLI+2009,ZINGER+2018}.

In all, the separation of clusters G5-7 from G1-4 in terms of both their galaxies' colours (i.e. those use as an input to the clustering, in particular their $NUV-u$ and $NUV-r$ colours) and morphologies (i.e. their higher S\'ersic indices), implies that their galaxies are subject to additional evolutionary processes. Hence, we suggest that the strong overlap between the passivity and the morphologies of G5-7 galaxies appears to be a product of different sets of environmental processes, which drive their quenching and morphological transformation separately \citep{POGGIANTI+1999,KELKAR+2019}. In addition, it implies that the quenching of galaxies precedes, or at least be simultaneous to, their morphological transformation \citep{SCHAWINSKI+2014,WOO+2017}. While the merger of two gas-rich, star-forming galaxies may produce a rejuvenated remnant, mergers between passive progenitors will invariably produce passive remnants with increasingly spheroidal morphologies, ranging from lenticular galaxies with classical bulges \citep{MISHRA+2017,MISHRA+2018,MISHRA+2019} through to pure spheroids.

\begin{figure*}
\centering
\includegraphics[width=0.78\textwidth]{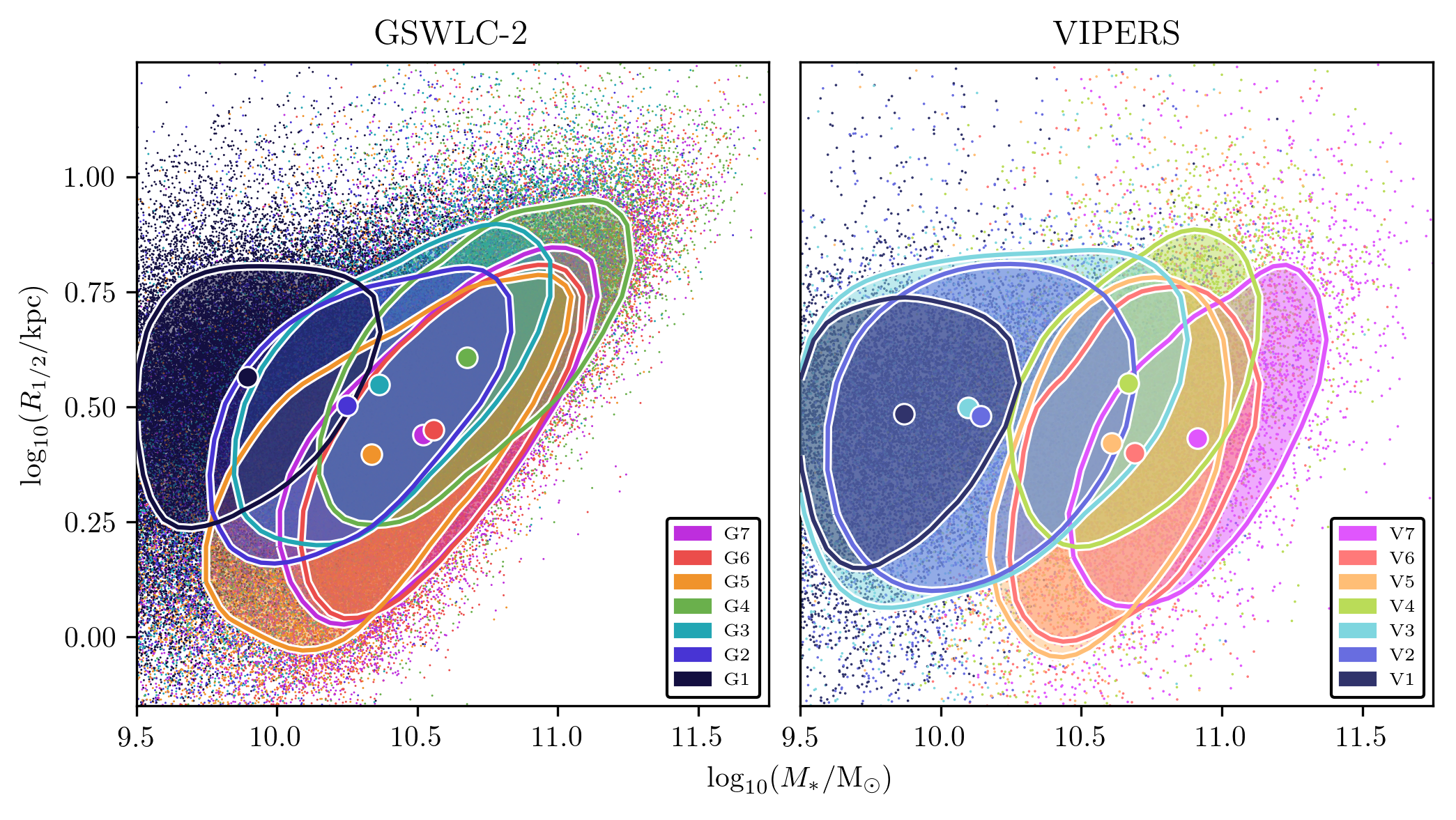}
\caption{Half-light radius versus stellar mass for the galaxies in our samples. Circularised half-light radii are calculated from single S\'ersic fits by \protect\cite{SIMARD+2011} for our GSWLC-2 sample, and \protect\cite{KRYWULT+2017} for our VIPERS sample. The distributions of clusters are shown using coloured, filled contours (drawn at a relative density of $0.4$), and the coloured, circular markers show their medians.}
\label{fig:massre}
\end{figure*}

\subsection{Clusters in the size-mass plane}
\label{subsec:massre}

Fig. \ref{fig:massre} shows the size-mass distribution of the clusters in each of our samples. The stellar masses originate from the same \texttt{CIGALE} SEDs that were used to generate the colours with which we represent the galaxies for the clustering, and the half-light radii from fits of single S\'ersic profiles (see Sections \ref{subsubsec:finalg} and \ref{subsubsec:finalv}). The size of a galaxy, in the context of its stellar mass and its morphology, is another important record of its assembly history. The positions and distributions of both sets of clusters in these plots match well with broader blue versus red, and early- versus late-type distinctions made in the same (or similar) plane(s) by other studies \citep{SHEN+2003,VANDERWEL+2014,LANGE+2015}. This result again demonstrates that \texttt{FEM}, via just the nine input colours, is able to identify subpopulations that are degenerate in two dimensions and that are ordinarily distinguished using a combination of photometric and morphological information.

The most significant difference between the two plots in Fig. \ref{fig:massre} is the absence of compact massive galaxies in our GSWLC-2 sample in comparison with our VIPERS sample. The canonical explanation for the growth of these galaxies is ongoing minor merger activity and accretion \citep{NAAB+2009,HOPKINS+2010}. The resultant shift between the passive VIPERS clusters and the passive GSWLC-2 clusters is approximately in accordance with the expected redshift evolution of the size-mass relation for early-type, passive galaxies \citep{VANDOKKUM+2015}, though the mass-incompleteness of passive VIPERS galaxies means that we are unlikely to have precisely captured this shift in this paper. The large overlap of G4 and V4 with their respective passive clusters in Fig. \ref{fig:massre} seems to support the additional `late-track' (late with respect to cosmic time rather than to morphology) of galaxy evolution proposed by \cite{BARRO+2013} to yield disc-dominated passive galaxies \citep{ILBERT+2010,CAROLLO+2013,SCHAWINSKI+2014}. Both sets of SFMS clusters are similarly distributed, capturing the minimal evolution of the sizes of star-forming galaxies between their two redshifts \citep{LILLY+1998,VANDERWEL+2014}.

\section{Summary and Conclusions}
\label{sec:conc}
 
We present results from the application of the \texttt{FEM} clustering algorithm to samples of galaxies at low ($z \sim 0.06$, from GSWLC-2) and intermediate ($z \sim 0.65$, from VIPERS) redshifts. Galaxies are represented using nine UV-through-NIR broadband rest-frame colours, derived from fits of ensembles of synthetic spectra to observed photometry with \texttt{CIGALE}. Using unsupervised machine learning to characterise the structures of our samples in this nine-dimensional feature space, our aims (following \citealt{SIUDEK+2018a}) were to  understand the evolution of subpopulations of galaxies in terms of these colours over cosmic time, and to establish how strongly these colours alone encode the assembly histories of galaxies. An advantage of \texttt{FEM} is its incorporation of dimensionality reduction on the fly, which ensures that it determines clusters using only the most important and discriminative information available among the input features. We summarise our results as follows:

\begin{enumerate}

\item Our cluster evaluation search reveals that both of our samples are best partitioned into seven clusters (Table \ref{tab:icl}). In addition, the best-fitting submodels to each of our samples, identified independently, are closely related, both allowing variation in the shapes of clusters and differing only in their treatment of `noise' among the input features. For both samples, these seven clusters break down into four `blue' clusters containing mostly star-forming galaxies (and the vast majority of green valley galaxies), and three `red' clusters containing mostly passive galaxies (Fig. \ref{fig:nuvrks}). These two families of clusters are clearly separable, both in terms of the input colours to the clustering as well as in terms of ancillary features, which suggests differences in the evolution of their galaxies. Clustering outcomes in general are highly robust and reproducible.

\item Overall, \texttt{FEM} uses the nine rest-frame colours similarly to determine the partitions (Fig. \ref{fig:imp}), reducing the dimensionality of the feature space to $6$ in both cases. Altogether, optical colours are most important to the clustering; individually, UV colours are. The availability of photometry with which to constrain the SEDs of galaxies is advantageous to the clustering. UV colours are slightly more important to the clustering in our GSWLC-2 sample, which has more GALEX coverage than our VIPERS sample. Similarly, the lack of any NIR coverage for our GSWLC-2 sample means that NIR colours are less important to its clustering. However, given the broader overall similarity between the clustering structures of the samples (Fig. \ref{fig:sub}), it appears that clustering (a statistical method) combined with SED estimation (which can infer rest-frame magnitudes from incomplete photometry) has enabled us to partially `fill the gaps' of missing data in our samples.

\item Blue clusters (containing mostly star-forming galaxies and the vast majority of green valley galaxies) in both samples form clear morphological sequences (Fig. \ref{fig:massng}). The correlation between their median S\'ersic indices and their median stellar masses captures the growth of the bulges of their galaxies along the SFMS (Fig. \ref{fig:bt}). At the highest masses, this growth corresponds with a drop in specific star formation rates. Hence, the quenching of high-mass galaxies is influenced by their inner stellar densities, above a certain threshold, which appears to be linked with nuclear activity (Fig. \ref{fig:agn}). The retention of discs by the highest-mass galaxies along this morphological sequence indicates that some galaxies quench without fully transforming their morphologies. The lack of a strong trend of these clusters with local environmental overdensity (Fig. \ref{fig:env}) suggests that this evolutionary pathway is dominated by internal processes. This pathway, prominent at the epochs of both samples, appears consistent with `mass quenching', as proposed by \cite{PENG+2010}. In addition, the SFMS appears to be a two-dimensional projection of this pathway which, in nine dimensions, extends all of the way to high-mass passive galaxies that retain their discs. We expect that the long timescales involved would ultimately lead the VIPERS star-forming clusters to resemble the GSWLC-2 star-forming clusters by the present day.

\item Red clusters (containing mostly passive galaxies) are clearly separate from their corresponding sequences of blue clusters. Galaxies in red clusters in both samples have uniformly high S\'ersic indices, indicating a fundamental link between centrally-concentrated morphologies and passiveness (Fig. \ref{fig:massng}). Passive clusters in our low-redshift sample are separated from their respective sequence of star-forming clusters, particularly towards lower stellar masses (Figs. \ref{fig:massng} and \ref{fig:bt}). We assume that this separation originates from the influence of alternative or additional processes to those that dictate the evolution of actively star-forming galaxies. Invoking the offset of these low-redshift passive clusters to high local environmental overdensities (Fig. \ref{fig:env}), we suggest that some of their galaxies are satellites, and subject to external processes. The homogeneity of their early-type morphologies implies that their quenching precedes, or is at least simultaneous to, their morphological transformation. In all, this pathway appears consistent with `environment quenching' \citep{PENG+2010}. This morphological separation is not as apparent for the passive clusters in our VIPERS sample (Fig. \ref{fig:massng}), which is mainly due to incompleteness of low-mass passive galaxies (which would also be expected to trace high-density environments). Hence, we are prohibited from commenting on the prevalence of this evolutionary pathway at intermediate redshifts.

\end{enumerate}

Our study appears to confirm the existence of two distinct evolutionary pathways of galaxies through the green valley \citep{POGGIANTI+1999,FABER+2007,PENG+2010,BARRO+2013,FRITZ+2014,SCHAWINSKI+2014,MOUTARD+2016b}. We re-emphasise that while much of our interpretation involves the use of ancillary features (and especially morphological information), the separation of the clusters into two main families of blue/green and red clusters originates in the colours used as inputs to the clustering. Hence, these pathways appear to be strongly encoded within the SEDs of galaxies. Our results invite further investigation into the extent to which a galaxy's assembly history may be discerned purely from its SED.

The use of further ancillary features would be instrumental in further substantiating and constraining these pathways. A wealth of such features are available for our GSWLC-2 sample, due to its basis in SDSS. Examples include Galaxy Zoo 2 morphologies \citep{WILLETT+2013} which include bar and merger classifications, and \cite{YANG+2007} group memberships to enable a distinction between central and satellite galaxies. A more detailed analysis of our low-redshift sample in this manner is reserved for a future study. We note that the Galaxy And Mass Assembly project \citep{DRIVER+2009} could provide an alternative low-redshift sample, given its panchromatic data release \citep{DRIVER+2016} and its rich library of value-added catalogues \citep{BALDRY+2018}. The upcoming Deep Extragalactic VIsible Legacy Survey (DEVILS; \citealt{DAVIES+2018}), which aims to improve completeness at $0.3<z<1.0$, could be the basis for an improved intermediate-redshift sample upon its completion. Furthermore, the Legacy Survey of Space and Time \citep{IVEZIC+2019}, which will provide galaxy colours and morphologies together, constitutes a particularly promising foundation for a future follow-up study.

The incompleteness of low-mass passive galaxies at intermediate redshifts would be alleviated by moving to deeper surveys such as G10-COSMOS \citep{ANDREWS+2017} and 3D-HST \citep{MOMCHEVA+2016}, both of which also have panchromatic photometric data releases. This would enable an examination of environment quenching at earlier epochs, and of its proposed increase in prevalence at lower redshifts \citep{FOSSATI+2017,MOUTARD+2018,PAPOVICH+2018}. Surveys like this could also extend our comparison to redshifts as high as $z \sim 2$, thus facilitating the constraint of the changing balance of evolutionary pathways, informed by clustering of rest-frame colours, over a greater extent of cosmic time.

\section*{Data Availability}

The data underlying this article will be made available upon reasonable request to the corresponding authors.

\section*{Acknowledgements}

We thank Steven Bamford, Olga Cucciati, Marie Martig, Tsutomu T. Takeuchi, and our referee for feedback which improved the quality of our manuscript.

The work presented in this paper was conducted using the following software: the \texttt{scikit-learn} \citep{PEDREGOSA+2011}, \texttt{matplotlib} \citep{HUNTER-2007}, \texttt{scipy} \citep{JONES+2001}, and \texttt{numpy} \citep{OLIPHANT-2006,HARRIS+2020} packages for the \texttt{Python 3} programming language (\url{https://www.python.org}); the \texttt{Fisher-EM} subspace clustering package (\citealt{BOUVEYRON+2012}; known in this paper as \texttt{FEM}) for the \texttt{R} statistical computing environment \citep{R}; and the Starlink Tool for OPerations on Catalogues And Tables (\texttt{TOPCAT}; \citealt{TAYLOR-2005}). 

The construction of the GALEX-SDSS-WISE Legacy Catalogue (GSWLC) was funded through NASA award NNX12AE06G. 

The VIMOS Public Extragalactic Redshift Survey (VIPERS) was performed using the ESO Very Large Telescope, under the `Large Programme' 182.A-0886. The participating institutions and funding agencies are listed at \url{http://vipers.inaf.it}. 

Funding for SDSS-III was provided by the Alfred P. Sloan Foundation, the Participating Institutions, the National Science Foundation, and the U.S. Department of Energy Office of Science. The SDSS-III website is \url{http://www.sdss3.org}. 

SDSS-III is managed by the Astrophysical Research Consortium for the Participating Institutions of the SDSS-III Collaboration including the University of Arizona, the Brazilian Participation Group, Brookhaven National Laboratory, Carnegie Mellon University, University of Florida, the French Participation Group, the German Participation Group, Harvard University, the Instituto de Astrofisica de Canarias, the Michigan State/Notre Dame/JINA Participation Group, Johns Hopkins University, Lawrence Berkeley National Laboratory, Max Planck Institute for Astrophysics, Max Planck Institute for Extraterrestrial Physics, New Mexico State University, New York University, Ohio State University, Pennsylvania State University, University of Portsmouth, Princeton University, the Spanish Participation Group, University of Tokyo, University of Utah, Vanderbilt University, University of Virginia, University of Washington, and Yale University. 

ST has been supported by a United Kingdom Science and Technology Facilities Council postgraduate studentship. MS has been supported by the European Union's Horizon 2020 Research and Innovation programme under the Maria Skłodowska-Curie grant agreement (No. 754510), the Polish National Science Centre (UMO-2016/23/N/ST9/02963), and the Spanish Ministry of Science and Innovation through the Juan de la Cierva-formacion programme (FJC2018-038792-I). KM has been supported by the Polish National Science Centre (UMO-2018/30/E/ST9/00082). This research has been supported by the Polish National Science Centre (UMO-2018/30/M/ST9/00757) and the Polish Ministry of Science and Higher Education (DIR/WK/2018/12).


\bibliographystyle{mnras}
\bibliography{bib} 


\appendix

\section{Iterations of \texttt{FEM}}
\label{app:iter}

\begin{figure}
\centering
\includegraphics[width=0.45\textwidth]{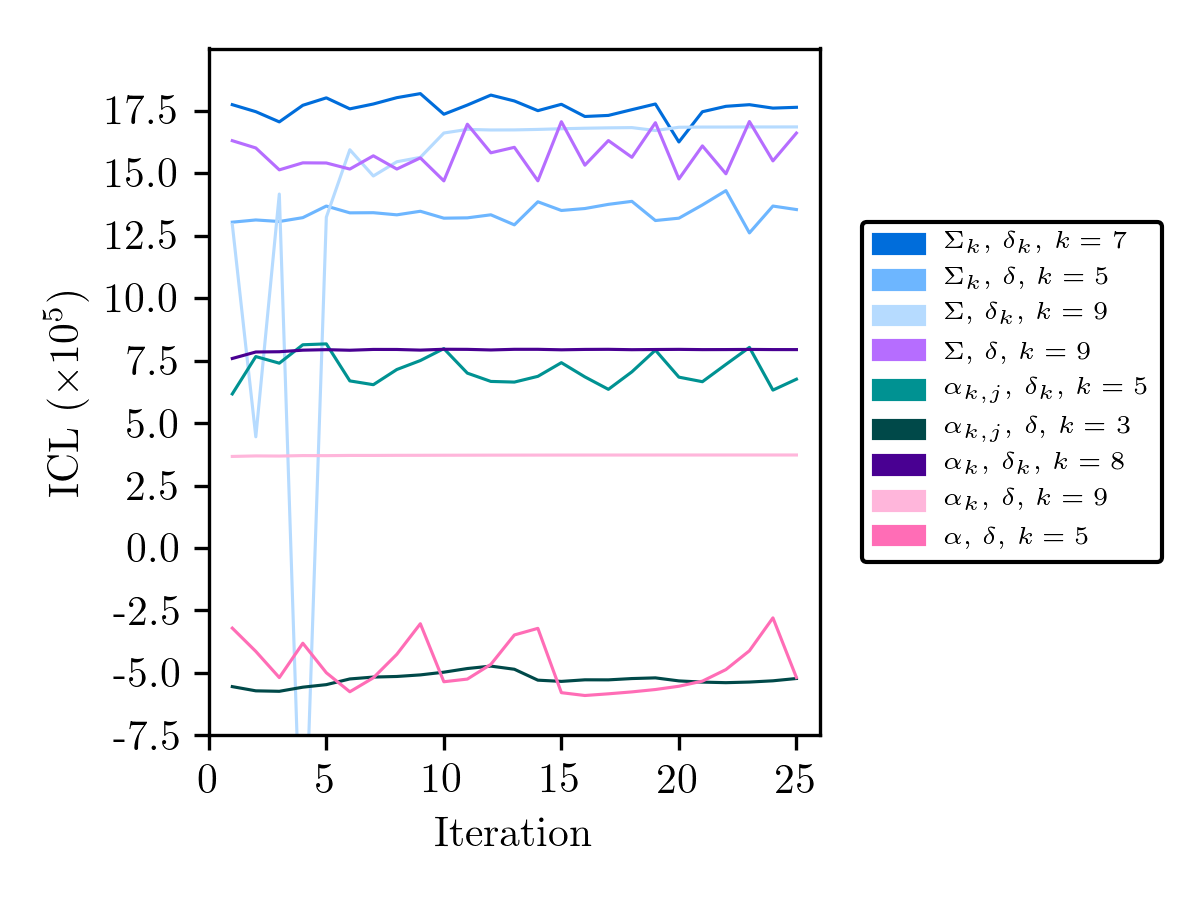}
\caption{ICL scores reported at iterations $1$ through $25$ by various combinations of submodel and $k$ for our GSWLC-2 sample. For each submodel, we show the value of $k$ which yields the highest ICL score. These iteration profiles are generally quite flat, indicating that \texttt{FEM} quickly converges to a stable outcome. The large changes exhibited by $\Sigma$, $\delta_{k}$, $k=9$ are due to the emptying of clusters as it iterates.}
\label{fig:iter}
\end{figure}

In Fig. \ref{fig:iter}, we show ICL scores reported at each of up to $25$ iterations by various combinations of submodel and $k$ for our GSWLC-2 sample. These `iteration profiles' are mostly quite flat; hence, $25$ iterations are more than sufficient for allowing \texttt{FEM} to stabilise to an outcome. In addition, the bulk of the clustering structure appears to be determined during the \texttt{k-means} initialisation step, which spreads the cluster centres out ahead of the first iteration. The ICL criterion rewards separated clusters, so \texttt{k-means} initialisations are particularly well suited to yielding useful clustering outcomes. Trials of the use of uniform random initialisations resulted in more combinations of submodels and $k$ failing to converge. 

Variations in the ICL values reported by individual combinations of submodel and $k$ over successive iterations arise due to the Fisher step of \texttt{FEM}, in which the subspace within which the clusters are to be modelled is found. Hence, the updating of the model parameters during the Maximisation step is \textit{indirectly} related to the probabilities calculated in the Expectation step. For traditional EM algorithms, these steps are directly related and thereby guarantee convergence. The large changes between successive iterations exhibited by some combinations (e.g. $\Sigma$, $\delta_{k}$, $k=9$) are most often due to the emptying of clusters; a reduction in the number of clusters used by \texttt{FEM} leads, in these cases, to a sudden increase in ICL.

\section{Smoothing of feature data for our GSWLC-2 sample}
\label{app:sm}

\begin{figure}
\centering
\includegraphics[width=0.44\textwidth]{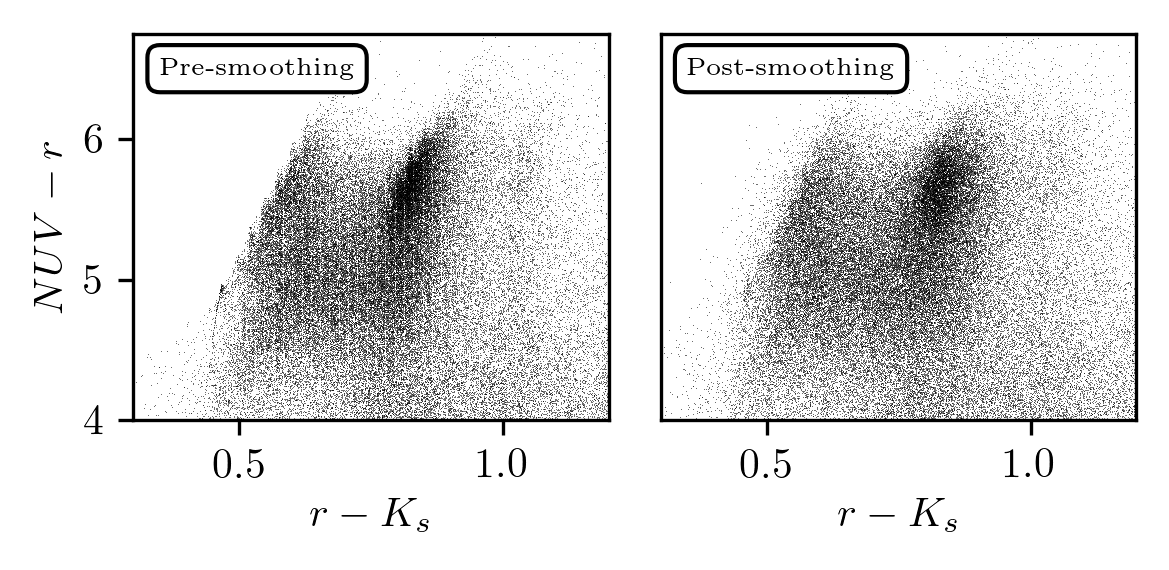}
\caption{The effect of our smoothing on the distribution of GSWLC-2 galaxies in the passive region of the $NUV-r-K_{s}$ colour-colour plane. Substructures in the distribution of galaxies within this region are preserved post-smoothing.}
\label{fig:sm}
\end{figure}

Preliminary tests revealed that a truncated, bimodal substructure among passive galaxies within the nine-dimensional colour space representing our GSWLC-2 sample (see the left-hand plot of Fig. \ref{fig:sm}; also visible in Fig. \ref{fig:sub}) led to an inability of \texttt{FEM} to converge for the majority of submodels and values of $k$. This truncated bimodal substructure is due to the lack of input NIR photometry to the \texttt{CIGALE} SED estimation of GSWLC-2 galaxies, such that their NIR SEDs must be inferred from UV and optical photometry. This, in turn, leads to poorly constrained, discretised metallicities: galaxies at $r-K_{s} \lesssim 0.67$  peak strongly at $\log_{10}(Z) \sim -2.4$, and those at $r-K_{s} \gtrsim 0.67$ at $\log_{10}(Z) \sim -2.1$. The NIR SEDs of VIPERS galaxies, on the other hand, are constrained by $K_{s}$-band photometry and hence have slightly more freedom to vary. This smooths their colour and metallicity distributions.

We hence opt to apply a small level of Gaussian smoothing to the GSWLC-2 distributions of the rest-frame absolute magnitudes reported by CIGALE. The smoothing scale for the rest-frame absolute magnitude of a given galaxy is given by its Bayesian error. These errors are winsorised at the mean value of the logarithmic distribution of errors (i.e. errors larger than the mean value are set to the mean value). This winsorisation ensures that the smoothing scale is kept small enough to avoid the potential loss of astrophysically meaningful substructures, while still enabling \texttt{FEM} to converge more readily. The absolute rest-frame magnitude most affected by this smoothing is $FUV$, whose errors are winsorised at a maximum value of $0.25$ (all other magnitudes have a maximum error $<0.1$ after winsorisation). The right-hand plot of Fig. \ref{fig:sm} demonstrates the effect of our smoothing, showing that the bimodality in the colours of passive galaxies is retained post-smoothing. While this bimodality is likely to be an artefact, trends in the astrophysical features of galaxies between its peaks are still likely to be genuine (see also Section \ref{subsubsec:pass}).

\section{Behaviour of the various submodels of \texttt{FEM} for our samples}
\label{app:icl}

Our model selection approach considers ICL scores for $72$ different combinations of submodel and $k$ for \textit{each} of our samples. The comparison of these $72$ combinations is simplified greatly by the realisation that several submodels exhibit consistent patterns of behaviour across all values of $k$.

\texttt{FEM} is unable to converge to an outcome for several combinations of submodel and $k$. The most common diagnosis made by \texttt{FEM} in the case of non-convergence is that a cluster has become empty (i.e. that it no longer contains galaxies). Table \ref{tab:icl} shows that several submodels are unable to converge beyond a maximum value of $k$, suggesting a limit to their ability to properly partition the samples. Alternatively, submodels that converge at $k$, but fail to converge at $k-1$ and $k+1$ appear to be striking a `sweet spot' in terms of this ability. Different combinations are generally very consistent with respect to convergence, converging for either all or none of our $100$ initialisations.

Given their flexibility and their high levels of parametrisation, the $\Sigma_{k}$, $\delta_{k}$ and $\Sigma_{k}$, $\delta$ submodels offer the greatest promise among all of the \texttt{FEM} submodels for yielding detailed and astrophysically meaningful partitions of our samples. The outcomes they produce are similar; they exhibit near-identical trends in their ICL scores for $k=2$ through $k=5$ for our GSWLC-2 sample in Table \ref{tab:icl}. They differ only in their treatment of the noise terms, which appears to be a minor detail in comparison with their shared use of full, unique covariance matrices. Outcomes at higher values of $k$ generally consist of splits of clusters present in outcomes at lower values of $k$. 

Submodels featuring non-unique covariance matrices for the Gaussian density functions representing the clusters (i.e. submodels with $\Sigma$ and $\alpha$, such that they all have the same shape) consistently produce clusters with highly disparate sizes. Some clusters are large, containing $30$ to $60$ per cent of the galaxies in our samples each (and each often spanning both blue and red galaxies); others are empty or nearly empty, containing $\lesssim 1$ per cent of the galaxies in our samples each. Nearly-empty clusters appear to capture small, undesirable artefacts in the structure of our samples within their input feature spaces. While it is unclear why \texttt{FEM} registers a valid ICL score for these outcomes when they include empty clusters (often cited as a cause for the failure of \texttt{FEM}; see above), it is clear that these submodels are too crude to return more than a very broad partition of our samples, and that their outcomes are limited in their capacity for astrophysical interpretation. All of this is also true for the $\Sigma$, $\delta$ clustering outcome at $k=9$ for our GSWLC-2 sample, which achieved the highest ICL score in our model selection search despite including empty and nearly-empty clusters. For these reasons, we reject this outcome for analysis.

\begin{figure}
\centering
\includegraphics[width=0.41\textwidth]{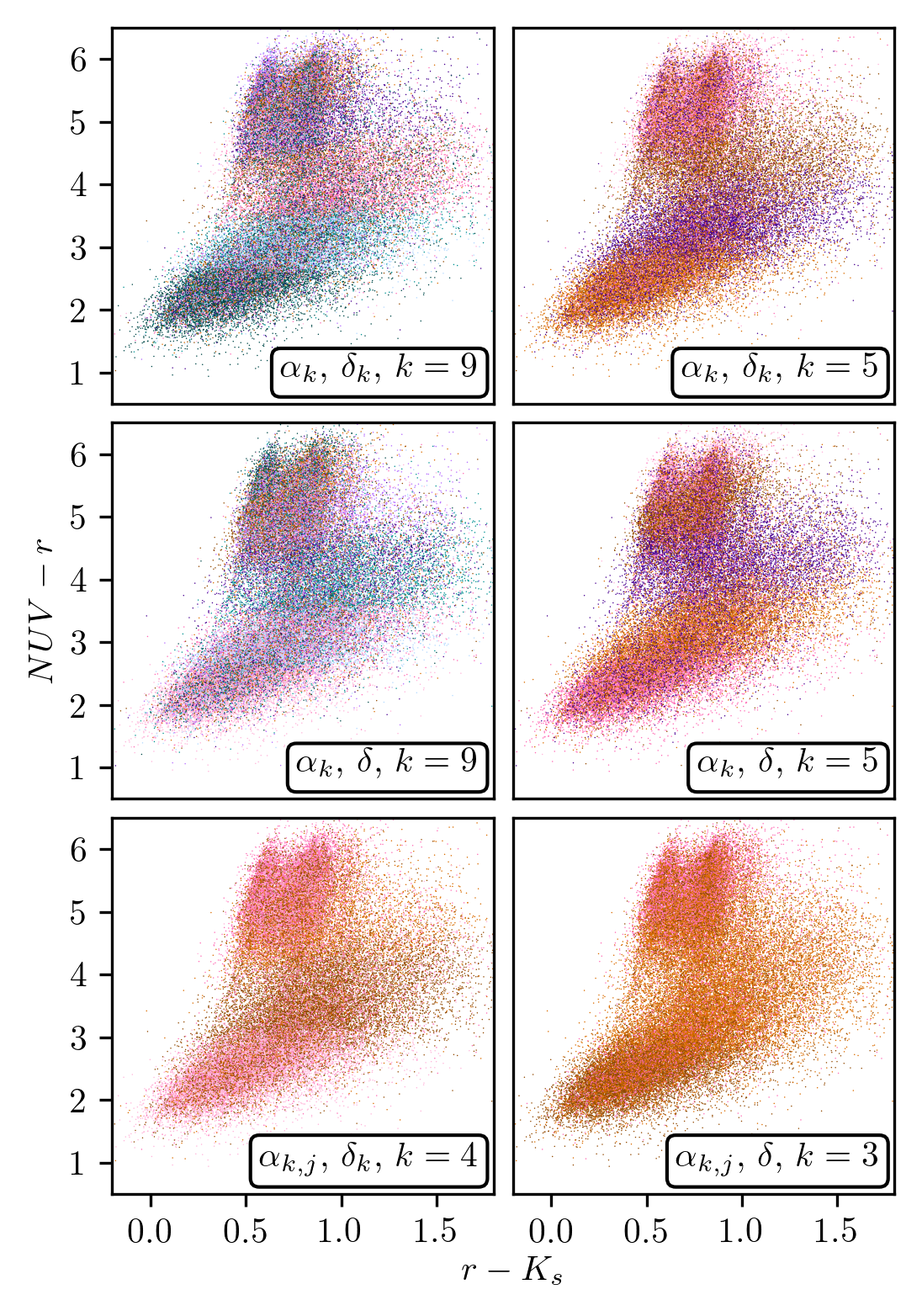}
\caption{Examples of the clustering structures determined by $\alpha_{k,j}$ and $\alpha_{k}$ submodels for our GSWLC-2 sample, shown in the $NUV-r-K_{s}$ colour-colour plane. The combination of submodel and $k$ for each outcome is shown to the lower-right of each plot. Individual galaxies are coloured in accordance with the cluster to which they belong. The choice of colours in this figure is not intended to imply any trends within or between plots. The horizontal striping pattern exhibited by these examples in these plots, which is a general property of $\alpha_{k,j}$- and $\alpha_{k}$-based outcomes, indicates segmentation mainly along a single axis.}
\label{fig:alpha}
\end{figure}

A general property of clustering outcomes reported by submodels which assume diagonal covariance matrices ($\alpha_{k,j}$, $\alpha_{k}$) for the Gaussian density functions within the discriminative latent subspace is that they segment our samples principally along a single dimension. Several representative examples of their clustering structures are shown in Fig. \ref{fig:alpha}, revealing that this single dimension is most strongly associated with the UV colours among our nine input features, with little-to-no distinction made between galaxies based on their NIR colours. We note that these submodels scored highest when we tested clustering of our samples using $i$-band magnitudes of galaxies as a reference point for defining colours (as in \citealt{SIUDEK+2018a}; see also Section \ref{subsec:feats}), producing the same striping pattern within the $NUV-r-K_{s}$ plane. While this simple segmentation does correspond broadly with incremental changes in the star formation activity of galaxies within our samples, other submodels (with $\Sigma_{k}$) return more detailed partitions and achieve higher ICL scores anyway.

The large spread in the ICL scores reported in Table \ref{tab:icl} arises directly from a large spread in the log-likelihood values of the fits. This large spread in the log-likelihood values arises, in turn, primarily from a $1 / \delta_{k}$ coefficient in the log-likelihood function of DLM model (which may be seen in full in appendix 2 of \citealt{BOUVEYRON+2012}). Submodels which yield very large but negative log-likelihood (and hence, ICL) values tend to have very small $\delta_{k}$ values for most (if not all) of their clusters; usually $0.001$, which is the floor that \texttt{FEM} imposes upon the value of $\delta_{k}$. Very small values of $\delta_{k}$ produce very large, positive values of $1 / \delta_{k}$, and (via a $- 1/2$ coefficient of the log-likelihood function) very large, negative values of the log-likelihood and, thus, of the ICL criterion. The addition of this especially low-variance noise to subspace Gaussians leads to highly peaked full space Gaussians which are unlikely to reflect the more continuous distributions of both samples (see Fig. \ref{fig:sub}).

\bsp
\label{lastpage}
\end{document}